\documentclass[twocolumn,times,tighten,twocolappendix]{aastex63}
\usepackage{graphicx,amssymb,amsmath,amsthm,amsfonts,epsfig}
\usepackage{mathtools}
\usepackage{amsbsy}
\usepackage{bm}
\usepackage{float}
\newcommand{\RA}{R_{\text{A}}}

\newcommand{\be}{\begin{equation}}
\newcommand{\ee}{\end{equation}}
\newcommand{\nn}{\nonumber}
\newcommand{\Medd}{\dot{M}_{\text{Edd}}}

\newcommand{\n}{\hat{n}}

\newcommand{\Rs}{R_{\star}}
\newcommand{\Ms}{M_{\star}}

\newcommand{\m}{{\rm m}}

\newcommand{\rrm}{{\rm m}}
\newcommand{\rco}{{\rm co}}

\newcommand{\rA}{{\rm A}}
\newcommand{\rX}{{\rm X}}

\def\apj{{ApJ}}
\def\aj{{AJ}}
\def\apjs{{The Astrophysical Journal Supplement}}
\def\apjl{{ApJL}}

\def\aap{{A\&A}}

\def\mnras{{MNRAS}}

\def\nat{{Nature}}
\def\aapr{{The Astronomy and Astrophysics Review}}

\def\prd{{Physical Review D}}
\def\prl{{Phys. Rev. Lett.}}

\def\04a{{2004 a}}
\def\04b{{2004 b}}

\begin{document}

\title{Does the gamma-ray binary LS I +61\textdegree303 harbor a magnetar?}
\author{Arthur G. Suvorov} \thanks{arthursuvorov@manlyastrophysics.org} 
\affil{Theoretical Astrophysics, Eberhard Karls University of T{\"u}bingen, T{\"u}bingen, D-72076, Germany}
\affil{Manly Astrophysics, 15/41-42 East Esplanade, Manly, NSW 2095, Australia}

\author{Kostas Glampedakis} \thanks{kostas@um.es}
\affil{Departamento de F{\'i}sica, Universidad de Murcia, Murcia, E-30100, Spain}
\affil{Theoretical Astrophysics, Eberhard Karls University of T{\"u}bingen, T{\"u}bingen, D-72076, Germany}

\date{Accepted ?. Received ?; in original form ?}

%\pagerange{\pageref{firstpage}--\pageref{lastpage}} \pubyear{?}

\AuthorCallLimit=2
\label{firstpage}

%%%%%%%%%%
\begin{abstract}
\noindent The high-mass X-ray binary LS I +61{\textdegree}303 is also cataloged as a gamma-ray binary as a result of frequent outbursts at TeV photon energies. The system has released two soft-gamma flares in the past, suggesting a magnetar interpretation for the compact primary. This inference has recently gained significant traction following the discovery of transient radio pulses, detected in some orbital phases from the system, as the measured rotation and tentative spin-down rates imply a polar magnetic field strength of $B_p \gtrsim 10^{14}\,\mbox{G}$ if the star is decelerating via magnetic dipole braking. In this paper, we scrutinize magnetic field estimates for the primary in LS I +61{\textdegree}303 by analyzing the compatibility of available data with the system's accretion dynamics, spin evolution, age limits, gamma-ray emissions, and radio pulsar activation. We find that the neutron star's age and spin evolution are theoretically difficult to reconcile unless a strong propeller torque is in operation. This torque could be responsible for the bulk of even the maximum allowed spin-down, potentially weakening the inferred magnetic field by more than an order of magnitude.
\end{abstract}
%%%%%%%%%


%%%%%%%%%%%%%%%%%%%%%%%%%%%%%%%%%

\section{Introduction} 
\label{sec:intro}
 
\cite{weng22} recently detected radio pulsations from the gamma-ray binary LS I +61{\textdegree}303 (henceforth LSI) using the Five-hundred-meter Aperture 
Spherical radio Telescope (FAST). Barycentric correction and pulse alignment established a millisecond rotation period for the primary, 
$P = 269.15508(16)$ ms, together with a spindown rate of $\dot{P} = 4.2(1.2) \times 10^{-10} \text{ ss}^{-1}$ 
{(though this measurement should be viewed as tentative; see Sec. \ref{sec:sec2}).}
The unambiguous detection of radio pulsations from LSI settles the long-held debate about the nature of the primary object as a neutron star, which was inconclusive 
from other, multi-wavelength data \cite[][and references therein]{tor10}. As a {couple} of soft-gamma flares have been detected in the direction of LSI \citep{dub08,bur12}, 
it is tempting to theorise that the primary is a magnetar \citep{tor12,pap12}. This proposition is now supported by magnetic-dipole braking theory which predicts a polar field 
strength of $B_{p} \approx 6.4 \times 10^{19} \sqrt{P \dot{P}}\text{ G} \sim 7 \times 10^{14}$~G, which ranks quite highly amongst the Galactic magnetar population\footnote{A list of known magnetars, together with their observed properties, is maintained at \url{http://www.physics.mcgill.ca/~pulsar/magnetar/main.html}.} \citep{mcg14}.

In this work, it is our goal to collect the various pieces of evidence from observations of LSI to re-examine the magnetar hypothesis. {In this context, we define `magnetar' specifically through an ultra-strong magnetic field, rather than using an empirical definition involving outburst activity.} One piece of information concerns 
viable scenarios for {explaining the $P$ and $\dot{P}$ values obtained from radio timing}. Although subject to (super-)orbital variability, likely owing to the high 
eccentricity of the binary [$e=0.63 \pm 0.11$ \citep{cas05}, though see \cite{krav20} who argue $e<0.2$], 
the X-ray luminosity in the 3--10~keV band is relatively modest, $10^{33} \lesssim L_{\rm X} / \mbox{erg s}^{-1} \lesssim 10^{34}$, as the system cycles between apastron (low) 
and periastron (high) \citep{rom07,esp07,had12}. This suggests a sub-Eddington accretion rate even at periastron, and therefore that a propeller torque may have a hard time 
contributing to the bulk of the spindown \citep{ghosh79b,wang95,accbook}. However, as recently shown by us in \cite{gs21}, large torques can be achieved when relaxing 
various assumptions about the magnetospheric geometry and the strength of induction-generated toroidal fields \cite[see also][]{watts22}. Using the models developed therein 
we show that the spindown data does not necessarily point towards a magnetar in LSI, even if we accept such a large $\dot{P}$.

Another consideration concerns the switch-on of the object as a radio pulsar. Conventional wisdom suggests that electron-positron pair production, occurring in 
unscreened magnetospheric `gaps', is a necessary ingredient to {excite radio pulsations in a neutron star magnetosphere} \citep{gj69,rs75} 
\cite[though see][for a critique]{melrose21}. 
Depending on the multipolarity of the {neutron star's magnetic field} and the structure of these gaps, a broadly-defined `death valley' separates inactive 
pulsars from those able to pair produce \cite[{see Sec. \ref{sec:deathval};}][]{cr93,ha01}. As LSI is rotating quite rapidly, the magnetic field required to avoid the death valley is moderate 
(at most $\sim 10^{13}$~G with an outer gap), and the radio activity of the source is unsurprising. {An obvious question then is why have radio pulsations only been observed now? This is discussed in Sec. \ref{sec:rpuls}.}

\cite{bed09b,bed09} \cite[see also][]{khang07,pap14} has argued that the gamma-ray shining of the binary mandates a minimum surface $B$-field strength of order $\sim 10^{14}$~G; the acceleration of charges near the magnetospheric boundary via Fermi processes, at the rate necessary to produce the $\lesssim$~TeV {radiation frequently observed from the source} \citep{alb06, acc08}, requires strong Poynting fluxes. This estimate, however, depends on the location of the Alfv{\'e}n surface, amongst other things, which is sensitive to the accretion geometry. {Another possibility involves colliding winds; see Sec. \ref{sec:gammaem}.}

The above considerations are discussed in this work, which is organised as follows. Section~\ref{sec:sec2} provides a detailed overview of the
observed properties of LSI, preparing the ground for the theoretical analysis of the subsequent sections. Section~\ref{sec:brake} forms the main part of the paper
and is devoted to the spin evolution and accretion dynamics of LSI. The physics related to gamma-band emissions from LSI are discussed
in Section~\ref{sec:tevgr}. The specifics of radio pulsar activation for LSI is the subject of Section~\ref{sec:deathval}.
Section~\ref{sec:GWs} is a digression to the paper's main topic, discussing the gravitational-wave (GW) observability of LSI by present and near-future detectors. Concluding remarks 
can be found in Section~\ref{sec:conclusions}. 

Throughout we use an asterisk label to denote stellar parameters like mass, radius, and moment of inertia. We also adopt the following
normalisations: $M_{1.4} = \Ms/1.4 M_\odot$, $R_6 = \Rs/10^6\,\mbox{cm}$, $\dot{M}_{n} = \dot{M}/10^{-n} M_\odot \mbox{yr}^{-1}$, $L_{n} = L_\rX /10^{n}\,\mbox{erg}\, \mbox{s}^{-1}$, $B_{n} = B/10^n\ \mbox{G} $, and $P_{-1} = P/0.1\,\mbox{s}$.

%%%%%%%%%%%%%%%%%%%%%%%%%%%%%%%%%%%%%%%

\section{An overview of the observed properties of LS I +61{\textdegree}303}
\label{sec:sec2}

\cite{weng22} detected 42 single pulses from LSI using FAST data from January 2020 which, after alignment and barycentric correction, 
revealed a source pulsating with period $P = 0.269$~s. The best-fit value for the period derivative
was found to be $\dot{P} = 4.2(1.2) \times 10^{-10} \text{ ss}^{-1}$. {This value, quoted from Supplementary Figure 1 
in \cite{weng22}, was obtained by folding data to maximise the signal-to-noise ratio using the 
\emph{prepfold} pipeline within the PulsaR Exploration and Search TOolkit (PRESTO) package \citep{ran02}. The cited period derivative and its uncertainties 
are therefore not obtained from direct timing. As noted by \cite{weng22}, there are only 3 hours worth of observations spanning a tight orbital 
phase $(\sim 0.58)$, and thus the authors were unable to recover the Doppler-shifted signals necessary to determine the intrinsic period derivatives. We can estimate the extent to which (orbitally-averaged) Doppler modulation imprints on the period derivative through $|(\Delta \dot{P})_{\rm Dop}| \leq P (2 \pi / P_{\rm orb})^2 a \sin i (1 + e \cos \omega)$ \cite[see, e.g.,][]{israel17}. Here, $a \sin i$ and $\omega$ are the projected semi-major axis in units of light-seconds and longitude of periastron, respectively. Under the most favourable orbital solution of \cite{cas05} (see Table 3 therein), we find $|(\Delta \dot{P})_{\rm Dop}| \lesssim 10^{-10}$, and thus $\dot{P}$ can be trusted to within a factor $\lesssim 2$. At worst, though, the maximum error could exceed the reported $\dot{P}$.}

Throughout this work, we operate under the assumption {of an upper limit $\dot{P} \leq 5.6 \times 10^{-10} \text{ ss}^{-1}$, and intend to show that, even if the extreme value is used, much of it could be attributed to accretion torques.}
In the remainder of this section we recap which observational arguments support either a magnetar or non-magnetar interpretation.

%%%%%%%%%%%%%%%%%%%%%%
\subsection{Radio pulsations: spindown}

A naive application of standard magnetic braking in vacuum implies a polar field strength of $B_{p} \lesssim 7 \times 10^{14}$~G, which is clearly of magnetar level. {The non-vacuum \cite{spit06} expression reduces this by only a factor $\lesssim 2$.} {A similar estimate for $B_p$ is obtained if we use the particle-wind modified formula 
of~\cite{harding99} and \cite{thomp00}, even if we assume a wind luminosity as high as the {maximum} spindown energy rate.}  As we show in detail in Sec. \ref{sec:brake} 
however, a strong propeller torque may be able to relieve the 
magnetic field from the spindown demands to some degree. For reasonable values of the torque particulars, we find that values of $B_{p} \sim 10^{13}$~G 
could be a viable alternative, {even assuming such a large $\dot{P}$.}

%%%%%%%%%%%%%%%%%%%%%%%%%
\subsection{Radio pulsations: death valley}
\label{sec:rpuls}

The existence of LSI as a radio pulsar 
implies that the object should
reside outside of the `graveyard' \citep{rs75}, which implies a minimum $B$-field strength. As we show in Sec. \ref{sec:deathval} however, for polar-gap configurations this minimum is only of order $\sim 10^{11}$~G, though could reach $\sim 10^{13}$~G for an outer-gap \citep{cr93,ha01}. 

Given that pulsations were only 
observed recently however \cite[January 7, 2020;][]{weng22}, one may argue that magnetic substructures atop the crust \cite[`starspots';][]{zhang07} or in the 
magnetosphere \cite[`twists';][]{belo09} may have only just developed or (Hall-)drifted into regions that are conducive to radio activity. {Magnetospheric twists, injected following quake activity (that may have sparked the soft-gamma flares in 2008 and 2012), can survive on long diffusion timescales \cite[$\gtrsim$ yrs;][]{parfey13} and temporarily reconfigure the geometry of the emission zone. Furthermore, many of the known `radio magnetars' experience accelerated spindown following flare activity. \cite{arch15} reported that the spin derivative of 1E 1048.1--5937 varied by a factor $\sim$5, starting $\sim$100 days after each of its outbursts, oscillating for $\sim$ years before stabilising. A magnetar interpretation may therefore simultaneously explain a large $\dot{P}$ at the time radio pulsations were observed and why the source switched on at all.} 

{Alternative explanations involve the mode of accretion or dynamics related to the companion.} The ram pressure of infalling material may have temporarily subsided, allowing for the source 
to activate as a radio pulsar, similar to what is thought to happen for the `swinging' pulsars PSR J1023+0038 and IGR J18245--2452 \citep{tam10,pap13}. {The absence of X-ray pulsations casts doubt on this interpretation however.} Another possibility is that the pulsar beam is most often directed through the wind from the companion, and is regularly quenched because the region is optically thick to free-free absorption \citep{chern20}. {\cite{zdz10} estimate the optical depth as $\tau_{\rm ff} \approx 5 \times 10^{3} (a / 3 \times 10^{12} \text{cm})^{-3}$, which is $\gg 1$ even at apastron if $e \lesssim 0.6$.} 

{Regardless, the role of the vast aperture available to FAST is unquestionable, as the mean pulsed signal was at the $\gtrsim \mu$Jy level \citep{weng22}. This means we cannot conclude that the source is only now radio-loud. Future observations will help to determine whether the source is always `on' but only visible at certain orbital cycles, which would favour the latter explanations above, or has since shut off, which would point towards dynamical phenomena in the magnetosphere.}

%%%%%%%%%%%%%%%%%%%%%%%
\subsection{X-ray/soft-gamma emissions: bursts}
\label{sec:softx}

Bursts with luminosities exceeding $10^{37} \text{ erg s}^{-1}$, likely though not definitely associated with LSI \cite[cf.][]{mun07}, were detected 
by the Swift Burst Alert Telescope in 2008 \citep{dub08} and again in 2012 \citep{bur12}. These short ($\lesssim 0.3\, \mbox{s}$) bursts are characteristic 
of those observed in magnetars \cite[e.g., 1E 2259+586, with polar field strength $\sim 10^{14}\, \mbox{G}$;][]{gav04}. \cite{tor12} suggest that the 
low intensity of the bursts favour an interpretation of a modest-$B$ magnetar with $B \sim 5 \times 10^{13}$~G, and that they are at variance with 
known type I X-ray bursts and are unlikely to be accretion-driven. Numerical simulations of magnetically-induced stresses in neutron stars suggest that local 
field strengths of order $\gtrsim 2 \times 10^{14}$~G are necessary to fracture the crust \citep{lan15}, which is a popular 
model for driving flares in soft-gamma repeaters \cite[e.g,][]{perna11}.

{If the neutron star occasionally accretes, a stipulation that is defensible because $\sim$ks-long X-ray bursts have been observed \citep{li11}, its magnetic field may be `buried'. Burial reduces the global dipole moment while tangling the field near the surface. Simulations suggest that strong $(\gtrsim 10^{14}$G) patches within accreted mountains are consistently generated even if the global field, controlling magnetospheric radii and spindown, is $\lesssim 10^{13}$G \citep{suvm20,fuji22}. Buried fields may persist on Ohmic ($\gtrsim 10^{5}$yr) timescales \citep{vig09}.}

%%%%%%%%%%%%%%%%%%%%%%%%%%%%%%%%%
\subsection{X-ray emissions: persistent, hard}
\label{sec:persx}
 
\cite{had12} found that the 3--10\,keV flux for LSI is of order 
$\gtrsim 4 \times 10^{-12}\, \mbox{ erg cm}^{-2} \mbox{s}^{-1}$ at apastron and $\lesssim 2 \times 10^{-11}\, \mbox{ erg cm}^{-2} \mbox{s}^{-1}$ 
at periastron. The recent \emph{Gaia} survey indicates a distance of $d_{\text{LSI}}= 2.65(9)$~kpc \citep{gaia21}, implying that the overall X-ray luminosity is of order $10^{33} \lesssim L_{\rm X} / \mbox{erg s}^{-1} \lesssim 10^{34}$. {\cite{esp07} cite a column density of $N_{\rm H} = 5.7(3) \times 10^{21} \text{cm}^{-2}$ while \cite{frail91} give $N_{\rm H} \lesssim  10^{22} \text{cm}^{-2}$, however, implying a factor $\gtrsim 2$ uncertainty in the unabsorbed luminosity.} These persistent, \emph{nonthermal} X-ray emissions could be accretion-powered, implying that $L_\rX$ can be converted into an estimate for the mass accretion rate $\dot{M}$ (see Sec. \ref{sec:accretion}). {An alternative reservoir for $L_\rX$ could be the colliding wind (see Sec. \ref{sec:gammaem}), in which case any inversions for the accretion rate from $L_{\rm X}$ will be overestimates.} It is important to note however that the relevant $L_{\rm X}$ -- at the time the source was radio-visible -- is unknown. \cite{weng22} detected pulses when the system had an orbital phase of $\sim$0.58 (see Table 1 therein), but given the uncertainties in the orbital modelling itself 
\cite[see][for a detailed discussion]{krav20}, it is non-trivial to deduce the orbital separation at the observation time.

%%%%%%%%%%%%%%%%%%%%%%%%%%%%%%%%%%
\subsection{X-ray (non-)emissions: persistent, soft}
\label{sec:therm}

Despite long-term monitoring campaigns \cite[e.g.,][]{paer07,esp07}, there is little evidence for a thermal component to LSI's persistent emissions. 
Isolated magnetars display relatively high surface luminosities \citep{mcg14}, thought to be provided by the secular decay of their internal 
magnetic fields \citep{td93,td96}. 
A conservative upper limit is set by $L_{\rm X,therm} \leq L_{\rm X} \sim 10^{34} \text{ erg s}^{-1}$; unfortunately, this is of little use to
the present analysis as it is comparable to the typical magnetar quiescent luminosity~\citep{mcg14}. Magnetothermal simulations carried out by \cite{anz22}
show that the surface luminosity of a heavy ($M \sim 1.8 M_{\odot}$) but highly magnetised $(B \lesssim 10^{14}\, \mbox{G})$ neutron star can drop below $10^{34} \text{ erg s}^{-1}$ even after $\sim$ centuries if the nucleon and hyperon direct Urca processes are active \cite[see their Figures 5 and B1 and also][]{anz22b}. As such, in order for thermal emission limits to have a true impact as concerns the nature of LSI, $L_{\rm X,therm}$ would need to be pushed downwards by (at least) an order of magnitude. Although a more precise statement can be made using spectral fits, it is likely that even fields of strength $\lesssim 10^{15}$~G cannot be definitively ruled out if fast Urca mechanisms operate \citep{page91}. 

\subsection{Gamma-ray emissions}
\label{sec:gammaem}

As argued by \cite{bed09b,bed09} \cite[see also][]{khang07}, surrounding the line where magnetic pressure balances the gravitational 
pressure of infalling matter (i.e., the Alfv{\'e}n surface), there exists a turbulent region where electrons are accelerated via Fermi processes, especially if the neutron star velocity exceeds the 
local speed of sound at the inner boundary of the corona \cite[supersonic propeller state;][]{pap12}. Synchrotron processes and inverse-Compton 
scattering of radiation from the Be star companion results in the electrons achieving a Lorentz factor of order 
$\gamma_{\text{max}} \approx 1.8 \times 10^{6} B_{14}^{5/14} \dot{M}_{10}^{-3/7}$ [see Eq.~(14) in \cite{bed09}]. Requiring that 
$\gamma_{\text{max}} \gtrsim 10^{6}$ demands a minimum $B$-field therefore. 

{An alternative location for the production of high-energy emissions  is in a colliding wind \cite[e.g.,][]{mara81}. The continuous outflow of relativistic particles 
from the pulsar encounters the stellar wind from the Be companion, resulting in a termination shock at the point where the pressures balance \citep{dub06}. 
Particles are scattered at the shock front and accelerated. Given the high maximum spindown luminosity of the primary, the pulsar wind is easily energetic enough to accommodate TeV emissions \citep{dub13}. Using smoothed-particle hydrodynamic (SPH) simulations in 3D, \cite{rom07} found that the geometry 
of the termination front does not match the morphology of the radio observations in LSI, though their simulations adopt wind luminosities of 
$\gtrsim 10^{36}\, \mbox{erg s}^{-1}$, far lower than the maximum spindown luminosity. \cite{tor12} separately argued that it is difficult to explain 
the anti-correlation between GeV and TeV emissions from LSI with a colliding wind. The external torques associated with this model may be minimal due to a high degree of spherical symmetry, thus making magnetic spindown the bulk contributor to $\dot{P}$. A magnetar conclusion for a large $\dot{P}$ would therefore be hard to escape.
}

\subsection{Age limits}
\label{sec:age}

The lack of an associated supernova remnant for LSI \citep{frail87} implies a likely age of at least a few kyr \citep{pap12}. {By contrast, if magnetic braking dominates over any accretion-related torques and the braking index of the source is held constant at $n=3$ over the lifetime of LSI, one predicts an 
age of $\gtrsim 10$~yr.} This is obviously excluded by the observations dating back to the late 1950s \citep{hard59}. It should be noted however that magnetars, {and pulsars more generally \citep{shaw22}}, exhibit noisy spindown. For example, the magnetar Swift J1834.9--0846 resides at the centre of supernova remnant W41 \citep{tian07}, strongly suggesting their association \citep{gran17}. An $n=3$ history for the former implies a star of age $\tau_{\text{sd}} \approx 4.9$~kyr \citep{mcg14}, though the true age of W41 was estimated by \cite{tian07} to be between $60-200\, \mbox{kyr}$, depending on expansion assumptions (e.g., extent of radiative cooling).

In any case, it is difficult to imagine a scenario where the magnetic field gets significantly \emph{stronger}  over time, and  therefore magnetic-braking `today' should occur at a similar or weaker rate than has occurred historically. An alternative explanation to avoid this age-related problem is that 
the object possesses a weaker field, though is presently spinning down at a high rate due to a large propeller torque. 

Based on its kinematic velocity relative to the Heart Nebula cluster IC 1805, 
\cite{mir04} argue that LSI may have been ejected from the young complex of 
massive stars from whence it came $\approx 1.7 \pm 0.7$~Myr ago. Such an age is not unusual for binaries involving neutron stars, though is virtually impossible to 
accommodate with a (present-day) sub-second magnetar scenario, {as spindown and field decay prevent old objects from being both fast and strongly magnetised simultaneously}. 

\subsection{Magnetars in binaries?}

\cite{kl19} suggest that magnetars in binaries should be rare. They argue that previous suggestions that some high-mass X-ray binaries (HMXBs), {and ultra-luminous sources in particular}, contain magnetars are problematic \cite[cf.][]{israel17,pop18}. {The most natural evolutionary scenario involves one member from a binary star (e.g., of class OB) undergoing core collapse and, eventually, leaving behind an X-ray binary with a magnetar primary. Large angular momentum reserves may be necessary however to entice dynamo activity in the protostar to generate fields exceeding $\sim 10^{15}\, \mbox{G}$ \citep{td93}. \cite{kl19} argue in this case that the supernova will be superluminous, likely destroying the companion and leaving only an isolated magnetar; cf. \cite{pop16,white22}. }

{If instead a magnetar captures a companion, such a process must occur within $\sim$Myr after birth, else the field is likely to have decayed sufficiently to depose the star of its magnetar status. Though the probability of a capture happening within this time is low, the issue of magnetars in binaries becomes one of semantics to some degree, as there are some `low-field' magnetars \cite[most notably SGR 0418+5729, which 
houses a surface field of only $\sim 6 \times 10^{12}$~G;][]{rip13}. Fields of this order could persist over $\gtrsim$Myr timescales without significant decay, 
depending on the electrical conductivity of the crust, which could be hampered by impurities \citep{pop18}. Field decay can also be stalled by 
plastic flows in a highly-magnetised crust, as such flows tend to oppose the existing electron fluid motions \citep{gl21}, thus preventing the formation of small-scale, tangled fields that are 
most at risk of Ohmic decay.} 

The following sections offer a more detailed discussion of the above topics. Each one concludes with a `final verdict'  as to the 
magnetar versus no-magnetar nature of LSI. 

%%%%%%%%%%%%%%%%%%%%%%%%%%%%%%%%%%%%%%%%
\section{Spin history}
\label{sec:brake}

In this section, we show how combinations of electromagnetic braking with various field geometries, together with accretion-induced braking torques, 
can be applied to investigate the spin history of the primary in LSI. 

\subsection{Electromagnetic spindown}
\label{sec:embr}

Neutron stars intrinsically decelerate through electromagnetic and gravitational torques, the relative impacts of which can be quantified in terms of a \emph{braking index}, $n$. For a simple, centered dipole, one has $n=3$, while for a general $\ell$-polar field one has $n = 2 \ell + 1$. Braking indices $n<3$ are also theoretically possible in cases where particle winds make up 
the bulk of the spindown torque \citep{mara81,thomp00}, or if the star is precessing and/or oblique \citep{mel97,pet19}. 

Including corrections induced by a force-free magnetosphere according to the \cite{spit06} formulae, the spin evolution of an inclined rotator may 
be approximately described by \cite[see also, e.g.,][]{mast11}
\be 
\label{eq:braking}
\dot{P} \approx \left(2 \pi \right)^{n-1} \frac {B_{p}^2 P^{2-n} \Rs^{3+n}} {6 I_\star c^n} (1 + k \sin^2 \alpha),
\ee
for inclination angle $\alpha$, magnetospheric factor $k \approx 1$, and moment of inertia $I_\star$. Throughout this work we adopt $I_\star \approx 0.38 M_{\star} \Rs^2$, 
as pertains to a Tolman-VII equation of state \citep{lp01} for stellar mass $M_{\star}$. Eq.~\eqref{eq:braking} can be used to infer $B_{p}$ as a function of $n$ and $\alpha$, given {values} of $P$ and $\dot{P}$. More generally however, Eq.~\eqref{eq:braking} can 
be solved for some (time-dependent) choices of $n$ and $B_{p}$, and one can infer the star's age. Late-time solutions to Eq.~\eqref{eq:braking} are relatively 
insensitive to the exact value of the initial birth period, $P(0) = P_{0}$. In particular, the spindown age of the object is approximated by 
$\tau_{\rm sd} \approx P/[(n-1)\dot{P}]$. Demanding $\tau_{\rm sd} > 60$~yr at minimum {\cite[owing to the object's discovery by][]{hard59}}, one then obtains $n \lesssim 1.4$, which in turn implies in fact that  
$B_{p} \approx 10^{12}$~G for an orthogonal rotator -- non-magnetar. The absence of a detectable supernova remnant however suggests that $\tau_{\rm sd} \gg 10^{3}$~yr and, unless the braking index\footnote{{In what follows, the \emph{braking index} here applies only to intrinsic torques, in the sense of equation \eqref{eq:braking}. Even if external torques modify the overall spin evolution, so that the true observational braking index is different from what we call $n$ in \eqref{eq:braking}, the same symbol is adopted for ease of presentation.}} is very close to unity, additional physics is needed to explain the object's behaviour.

%%%%%%%%%%%%%%%%%%%%%%%%%%%%%%%%%%%%%%%%%%%%
\subsection{Accretion torques}
\label{sec:accretion}
%It should be emphasised that t

{Although accretion discs are rare in HMXB systems, \cite{hk04} found with SPH simulations that a persistent disc comes to surround the neutron star in Be/X-ray binaries with periods $P_{\rm orb} \sim 24$~days, independently of the model specifics. We are therefore justified} in describing LSI as an accreting system with a thin disc threaded by the neutron star's magnetic field
\cite[cf.][]{ghosh79b, accbook}. 
This working hypothesis comes with a certain degree of approximation; 
the geometry of the actual accretion flow could depart from the well-ordered axisymmetric structure of a thin equatorial disc, especially if there is a strong 
quasi-spherical wind component. To some extent, some of the key parameters discussed below, like the magnetospheric radius and the 
$L_\rX (\dot{M})$ relation, are robust enough as to remain accurate even if the accretion flow would not take the form of a thin disc. 
On the other hand, if we were to consider the opposite limit of
spherical Bondi accretion, say, then the resulting torque would be negligible and, as we discuss below, it is difficult to 
explain the $P, \dot{P}$ data of LSI.

The thin disc model includes two basic lengthscales; the first one is the corotation radius $R_\rco = (G\Ms/4\pi^2\nu^2)^{1/3}$ 
which marks the radial distance where the local Keplerian angular frequency of the orbiting gas matches the stellar angular frequency.  
The magnetospheric radius
\be
\RA = \xi \frac{B_p^{4/7} \Rs^{12/7}}{(G\Ms)^{1/7} \dot{M}^{2/7}},
\label{RAdef}
\ee
is defined by the energy balance between the gas and the stellar poloidal magnetic field. Note that while expression \eqref{RAdef} implicitly 
assumes the field is a dipole, multipolar corrections are small as an $\ell$-pole decays as $r^{-\left(2 \ell +1\right)}$, and is 
therefore weak at radii $r \gg \Rs$ \cite[see Sec. 4.1 of][]{gs21}. At this radius the accretion flow becomes dominated by the magnetic field and the 
disc is effectively truncated.  The parameter $\xi$, typically taking values in the range $0.1$--$1$, is a useful phenomenological device for describing the 
complex (and highly uncertain) physics taking place at the boundary layer around $\RA$ \cite[see][for details]{gs21}. The recent, general-relativistic simulations 
conducted by \cite{watts22} show that expression \eqref{RAdef} adequately describes the magnetospheric radius in 
most cases, though with the general trend that the effective $\xi$ decreases as the dipole field strength increases (see Table 1 therein).

In addition to the above two lengthscales, the light cylinder radius, $R_{\rm lc} = c/2\pi \nu$ for speed of light $c$, is a key quantity for the pulsar mechanism. One finds that
\begin{align} \label{eq:radii}
\frac{\RA}{R_\rco} &\approx 289\, \xi \frac{R_6^{12/7} B_{14}^{4/7}}{M_{1.4}^{10/21} \dot{M}_{-10}^{2/7}  P^{2/3}_{-1}},
\\
\nn \\
\frac{\RA}{R_{\rm lc}} &\approx 22\, \xi \frac{R_6^{12/7} B_{14}^{4/7}}{M_{1.4}^{1/7} \dot{M}_{-10}^{2/7}  P_{-1}}. 
\end{align}
The $\RA > R_\rco$ regime, likely the relevant one for LSI, is the so-called propeller regime: the disc is truncated
so far away from the star that the inward motion of the gas is centrifugally inhibited by the much faster rotating magnetic lines. 
This arrangement results in a net spindown for the neutron star. 

An $\RA > R_{\rm lc}$ arrangement means that the magnetic field is strong enough (and the accretion rate is weak enough) 
as to provide the neutron star with a clean, gas-free magnetosphere, thereby allowing for radio emission as in ordinary pulsars. 
Based on the above estimate, LSI could sit on either side of the fence, especially if $\xi \ll 1$. Given its high orbital eccentricity, the transition from radio emission to (pulsational) radio silence could even take place during an orbital period provided $\dot{M}$ is sufficiently variable from apastron to periastron, as in the `flip-flop' model of  \cite{tor12}.

The standard formula  $L_\rX = G \Ms \dot{M} /\Rs$ for estimating $\dot{M}$ is not applicable to a propellering system, where instead
\be \label{LNew}
L_\rX \approx G \Ms \dot{M} / \RA.
\ee
{A simple inversion then leads to
\be \label{eq:mdotinv}
\dot{M}_{10} \approx 0.71\, \xi^{7/9} L_{33}^{7/9}  B_{12}^{4/9} R_6^{4/3} M_{1.4}^{-8/9}.
\ee
This expression, when combined with our estimate for the non-thermal luminosity $L_\rX$,
leads to an accretion rate range,
\be 
\label{eq:mdotrange}
10^{-4} \lesssim \frac{\dot{M}}{ \dot{M}_{\rm Edd}} \lesssim 10^{-2},
\ee
for $\xi \lesssim 0.1$ (see below) and $1 \lesssim B_{12} \lesssim 100$, where $ \dot{M}_{\rm Edd} \approx 1.5 \times 10^{-8}\, M_\odot \,\mbox{yr}^{-1} $ 
is the Eddington accretion rate.}

The most basic torque associated with the thin disc model comes from infalling matter at the lever-arm distance $R_\rA$,
\be
N_\rA =  \dot{M} (GM_\star \RA)^{1/2}.
\ee
This is easily modified to account for the relative rotation between the disc and the magnetic lines, viz.
\be
N_\rrm = N_\rA (1-\omega_\rA), 
\ee
where the so-called fastness parameter is defined as $\omega_{\rm A} = (\RA/R_\rco)^{3/2} $. This torque
allows for spin equilibrium, $N_\rrm=0$, as well as a propeller regime, $N_\rrm <0$. 

A more rigorous modelling of the disc-magnetic field coupling should account for the induced azimuthal magnetic field component; 
the outcome is an accretion torque which comprises $N_\rA$ and a radially integrated contribution from the entire disc~\citep{wang95}. 
This torque has been revised by us in~\cite{gs21} to include the boundary layer parameter $\xi$, with the result being
\be
N_{\rm disc} = \frac{1}{3} \xi^{-7/2} N_\rA \left  ( 1 + 3 \xi^{7/2} - 2 \omega_\rA \right ),
\label{Nnew}
\ee
{which combines with the torque(s) implied by \eqref{eq:braking}.}
Once the torque is obtained, the spin up/down rate of the neutron star is easily found with the help of
$\dot{\nu} = {N}/(2\pi I_\star)$.

\begin{figure}
\includegraphics[width=\columnwidth]{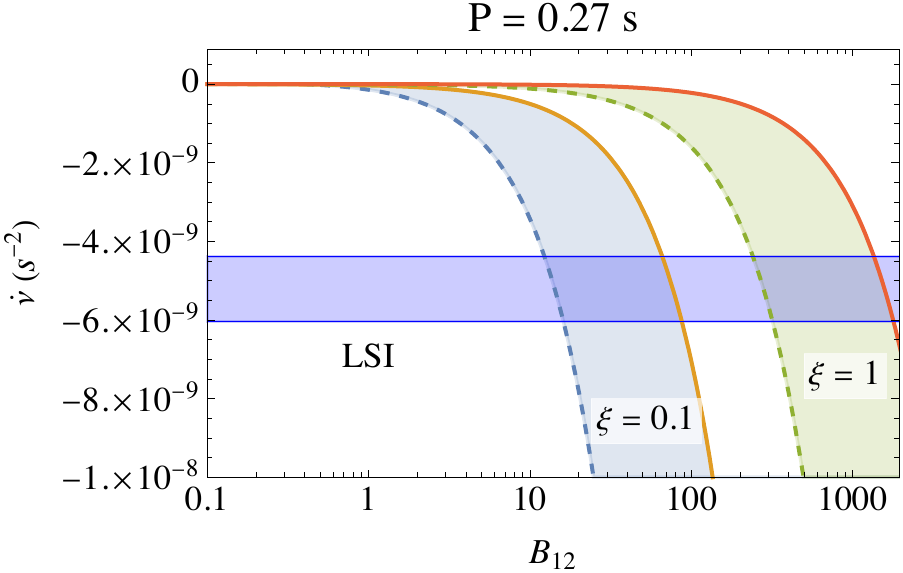}
\caption{The propeller spindown rate for LSI, $\dot{\nu}$, calculated with the torque~\eqref{Nnew}, as a function of the
polar field, $B_p/10^{12}\,\mbox{G}$.  Each of the two shaded bands corresponds to an $\dot{M}$ range, varying 
from $ \dot{M}= 10^{-2}  \dot{M}_{\rm Edd}$ (leftmost dashed curves) to $ \dot{M} = 10^{-4}  \dot{M}_{\rm Edd}$ (rightmost solid curves). 
{The former (latter) value corresponds to the estimated quiescent $L_\rX$ luminosity of LSI at periastron (apastron).} The left (right) band corresponds to $\xi=0.1$ ($\xi=1$).  The horizontal band marks the {maximum} spindown rate of LSI, viz.
$ - 6.04 < \dot{\nu}/10^{-9} \text{s}^{-2}  < - 4.39 $ \protect\citep{weng22}.}
\label{fig:propeller}
\end{figure}

The spindown rate associated with~\eqref{Nnew}
is shown in Fig.~\ref{fig:propeller}.  {Each band is delimited by a mass accretion rate specified by the range \eqref{eq:mdotrange}, where the lower limit (depicted with solid curves in Fig.~\ref{fig:propeller}) roughly corresponds to LSI's revised persistent X-ray luminosity at apastron, see Eq. \ref{LNew}. 
The dashed curves in Fig.~\ref{fig:propeller} correspond to periastron-like $\dot{M}$ values, more typical for low mass X-ray binaries.}
The $\xi =1$ spindown band is compatible with the {upper limits on $\dot{\nu}$} provided $B_p \gtrsim 5 \times 10^{14}\,\mbox{G}$. The $\xi=0.1$ 
band does the same job using a somewhat moderate field $10^{13}\,\mbox{G} \lesssim B_p \lesssim 10^{14}\,\mbox{G}$.
A value $\xi \ll 1$ is perhaps the most physically motivated; $\xi \lesssim 0.1$ is required in order to theoretically explain the observed spin up 
rates of several accretion-powered X-ray pulsars~\citep{gs21}, most notably SAX J1808.4--3658. {\cite{beck12} also argue for $\xi \sim 0.1$ in a number of systems from cyclotron line considerations.}

The above analysis suggests that LSI's tentative $\dot{\nu}$, or some other value within an order of magnitude, could be entirely or partially the result of propeller action, especially if 
the instantaneous $\dot{M}$ during the spindown measurement is a factor 10-100 higher than the estimated persistent
accretion rate. In that case the magnetic field inferred from the $P,\dot{P}$ data may be an overestimation of the neutron star's true magnetic field,
making LSI either a `low-$B$' magnetar or a `high-$B$' pulsar. 

%%%%%%%%
%%%%%%%%%

%%%%%%%%%%%%%%%%%%%%%%%%%
\subsection{Spin evolutions: weaker field}
\label{sec:weaker}

%%%%%%%%
\begin{figure*}
\includegraphics[width=2\columnwidth]{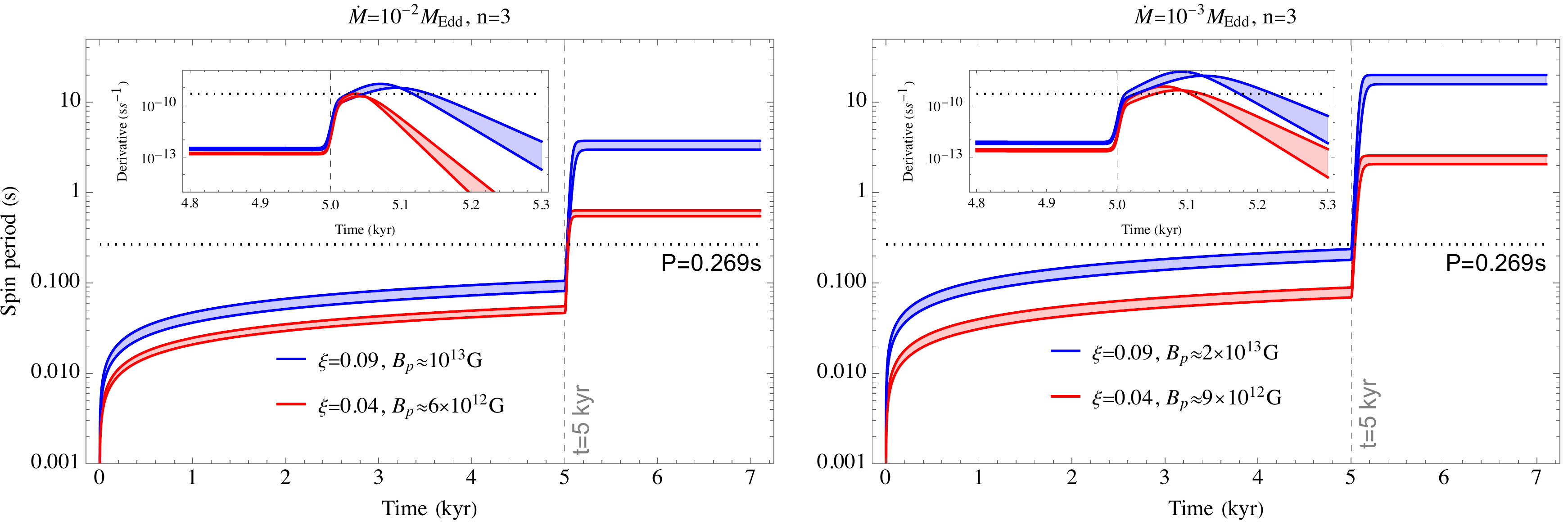}
\caption{`Weak field' $P(t)$ tracks for LSI, with assumed accretion rates of $\dot{M} = 10^{-2} \Medd$ and 
$\dot{M} = 10^{-3} \Medd$ for $t \geq 5$kyr in the left and right panels, respectively, though with negligible propeller torques at earlier times. 
The blue (red) curve, whose width accounts for a fiducial uncertainty in the {maximum} period derivative, viz. $3.2 \leq \dot{P}/10^{-10} \leq 4.4$,
has $\xi = 0.09 (0.04)$. We fix the (intrinsic) braking index at $n=3$ and inclination angle at $\alpha =0$. The polar field strength, as shown in the plot legends, 
is chosen such that the torques produce the requisite $\dot{P}$ when $P = 0.269$~s. The figure insets display the period derivative; the rapid rise occurs because strong propeller torques are turned on.}
\label{fig:Mdots}
\end{figure*}
%%%%%%%%%
%for times near $t = 5$~kyr;

As demonstrated in Sec. \ref{sec:accretion}, propeller torques alone with $B_{p} \gtrsim 10^{13}$~G can theoretically explain the {maximum} $\dot{P}$ value 
of the neutron star in LSI, provided that one uses the strong torque profile \eqref{Nnew} together with $\xi \sim 0.1$ and 
$\dot{M} \approx 10^{-2} \Medd$. However, the propeller torque will be aided by intrinsic braking torques, as described in Sec. \ref{sec:embr}. 

In general, if we wish to explain the  present-day $\dot{P}$ with a large $B$ field, the historical $\dot{P}$ is likely also to be large, implying a very young object. {This can be avoided if the  surface field is highly multipolar (see Sec. \ref{sec:stronger}), or if propeller torques activate sporadically, including just prior to when the radio pulses were observed (i.e., at a recent `switching time' defined below).} Figure \ref{fig:Mdots} shows a variety of possible evolutionary tracks, with the left (right) panel showing $\dot{M} = 10^{-2} (10^{-3}) \Medd$ with two different 
values of $\xi$ (blue and red curves), where propeller torques are considered small (i.e., $\dot{M} \ll \Medd$) at times $t \leq 5$~kyr, though become large 
at this fiducial transition point $t = 5$~kyr. At this `switching time', the star enters a phase 
of rapid spindown to allow for an enormous $\dot{P}$. This is illustrated within the figure inset, which shows a smooth though 
rapid increase in $\dot{P}$ at the switching time. In this simple picture, the stellar magnetic field is of order $B_{p} \lesssim 10^{13}$~G, 
as is typical for HMXBs \citep{pap12}. The switching time sets a floor to the age of the system ($\gtrsim$kyr), but also cannot be too large ($\lesssim$10kyr) for a 
given $B_{p}$, else the system will exceed the measured spin period without matching the $\dot{P}$ value. As can be seen in the $\dot{M} = 10^{-3} \Medd, \xi = 0.09$ 
case (right panel, blue curve), pure $n=3$ braking already has the star approach the present-day period 
value at $t \approx 5$~kyr. {The key points described above persist when involving multiple `switching times': if the system dances between propeller and accreting states, the number switching times could be large and the system could be old. Furthermore, the evolution should be `bumpier' than shown in reality, 
with periodic variations in $\dot{M}$ due to the orbital eccentricity.}

%%%%%%%%%%%%%%%%%%%%%%%%%
\subsection{Spin evolutions: stronger field}
\label{sec:stronger}

%%%%%%%%
\begin{figure*}
\includegraphics[width=2\columnwidth]{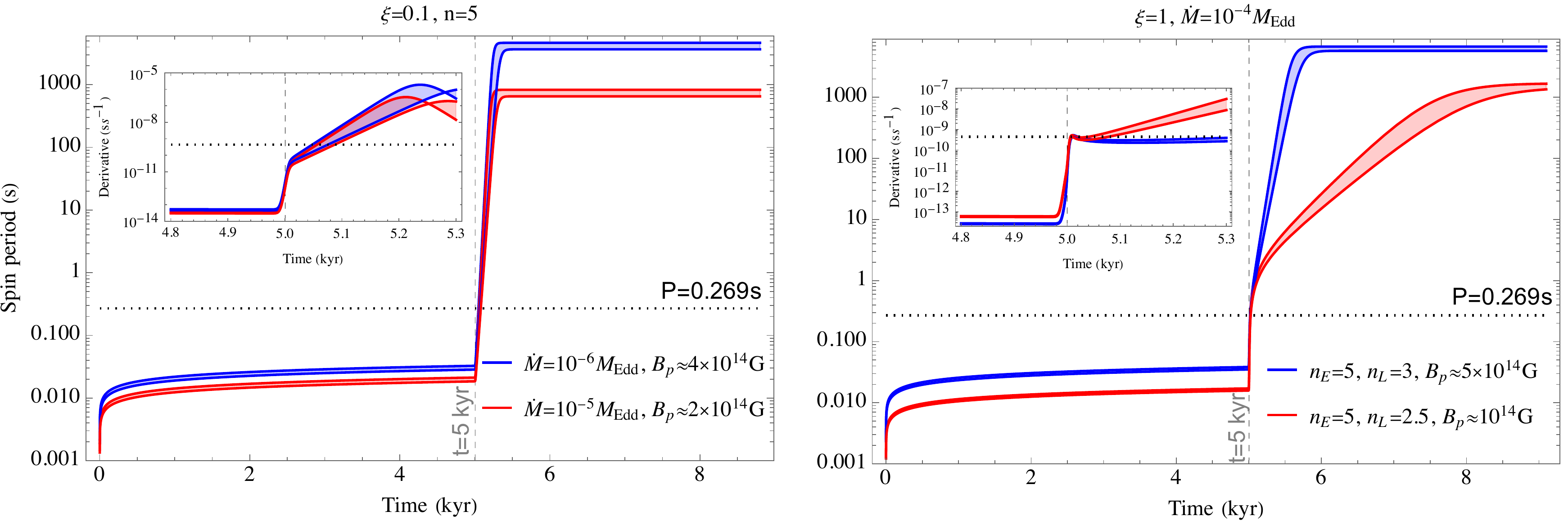}
\caption{`Strong field' spin period evolutions: $P(t)$ tracks for LSI permitting large polar field strengths, $B_{p} \gtrsim 10^{14}$~G, 
as shown in the plot legends. Larger braking indices allow for $\dot{P}$ to be relatively small at early times $t \leq 5$~kyr (see insets, which display $\dot{P}$ as in Fig. \ref{fig:Mdots}). Propeller torques may activate at this fiducial switching time for an accretion rate $10^{-6} \leq \dot{M}/\Medd \leq 10^{-5}$ (left panel), or the surface field may decay to a dipole-like configuration with $2.5 \leq n \leq 3$ (right panel), to account for {a large} $\dot{P}$ shortly after $t = 5$~kyr. We fix $\alpha = 0$.}
\label{fig:magfitsn5}
\end{figure*}
%%%%%%%%%

One way that a magnetar-like field can be accommodated, while also permitting a relatively old system, is if the historical braking index $n \gtrsim 5$. Such a value is appropriate for a quadrupolar field. Similar to Fig. \ref{fig:Mdots}, we suppose that the propeller mechanism activates at some fiducial time $t = 5$~kyr, with $\dot{M} = 10^{-5} \Medd$ (red curve) or 
$\dot{M} = 10^{-6} \Medd$ (blue curve), as shown in the left panel of Fig.~\ref{fig:magfitsn5}. Here we take $n=5$, which prevents magnetar-level polar field strengths, $B_{p} \gtrsim 10^{14}$~G, 
from slowing down the star too much at early times. {Alternatively, the magnetic geometry may have reconfigured near this switching time (see Sec. \ref{sec:rpuls}), such that the multipolarity of the field decreased.} The right panel shows an idealised situation where a pure quadrupole ($n=5$) reconfigures to a dipole-like ($n \approx 3$) field near the switching time instead, thus increasing the instantaneous spindown rate [see equation \eqref{eq:braking}]. Ignoring the Hall effect, a reduction in the field multipolarity as time progresses may be expected from Ohmic decay, as more tangled patches of field tend to dissipate quicker \cite[e.g.,][]{ag08}. 

{The conclusion of this section is that because it is unclear which (instantaneous) value of $\dot{M}$ applies for the 
period when \cite{weng22} timed the radio pulses (see Sec. \ref{sec:sec2}), several scenarios are theoretically possible. If the system is a magnetar (i.e., $B_{p} \gtrsim 10^{14}$~G), lower 
values of $\dot{M}$ (i.e., at apastron) in conjunction with a quadrupole-like braking index $n \sim 5$ can explain ages $t \geq 5$~kyr and the maximum $\dot{P}$ values (see Fig. \ref{fig:magfitsn5}). For a non-magnetar, a larger $\dot{M}$ (i.e., at periastron) combined with fields more typical of HMXBs could 
easily explain the observations with $B_{p} \lesssim 10^{13}$~G, if $3 \lesssim n \lesssim 5$ and $0.05 \lesssim \xi \lesssim 0.1$ (see Figs. \ref{fig:propeller} and \ref{fig:Mdots}). }
%If the true age of LSI exceeds a Myr, as anticipated by the source kinematics \citep{mir04}, it is most probable that the magnetic field has undergone significant decay \cite[as is believed to be the case for low-$B$ magnetars like SGR 0418+5729;][]{rip13} and the ensuing $\dot{P}$ must be lower than that reported by \cite{weng22}.

%%%%%%%%%%%%%%%%%%%%%%%%%%%%%%%%
\section{TeV gamma-rays}
\label{sec:tevgr}

LSI is one of the few known gamma-ray binaries, emitting $\sim$~TeV radiation, as confirmed by measurements taken with 
the MAGIC and VERITAS gamma-ray telescopes~\citep{alb06, acc08}. In this section, we consider a model of particle acceleration near the 
magnetospheric boundary, to estimate the minimum polar field strength of LSI, required to produce ultra-high energy radiation.

\subsection{Particle acceleration}

As put forth by \cite{bed09b,bed09} and others, electrons can be accelerated in the turbulent, transition region near the 
boundary of the Alfv{\'e}n surface through Fermi processes. The associated power, for an electron of energy $E$, can be parameterised as
\be
\label{eq:pacc}
W_{\text{acc}} = \zeta c E/R_{\text{Lar}} = \zeta c e B,
\ee
where $R_{\text{Lar}}$ is the Larmor radius, $e$ is the elementary charge, and $\zeta = 10^{-1} \zeta_{-1}$ is the dimensionless ``acceleration parameter'', which effectively sets a cutoff 
energy for the emitted spectrum; in the case of relativistic shocks, \cite{khang07} suggest that $10^{-4} \lesssim \zeta \lesssim 10^{-2}$, though this parameter is theoretically uncertain. We proceed by matching expression \eqref{eq:pacc} with the energy losses from synchrotron processes, which apply to the highest 
energy particles, viz.
\be
\label{eq:psic}
W_{\text{syn}} \approx  4 c \sigma_{T} \rho_{A} \gamma^{2} / 3,
\ee
where $\sigma_{T}$ is the Thomson cross-section and $\rho_{A} \approx B_{A}^2/8 \pi$ is the Alfv{\'e}n energy density in the transition region. 
From the definition of the magnetospheric radius \eqref{RAdef}, we find
\be
B_{A} \approx \frac {(G \Ms)^{3/7} \dot{M}^{6/7}} {(B_{p} \Rs^3)^{5/7} \xi^3}.
\ee
Maintaining all of the various scalings, we therefore obtain
\be \label{eq:gammamax}
\hspace{-0.2cm} \gamma_{\text{max}} \approx 3.9 \times 10^{6} \zeta_{-1}^{1/2} \xi^{3/2} B_{14}^{5/14} \dot{M}_{10}^{-3/7} M_{1.4}^{-3/14} R_{6}^{15/14}.
\ee
In general, the relativistic electron energy, $E = \gamma m_{e} c^2$ for mass $m_{e}$, acquires a $\sim$~TeV 
value for $\gamma_{\text{TEV}} \sim 2 \times 10^{6}$. Demanding that $\gamma_{\text{max}} \gtrsim \gamma_{\text{TEV}}$, we obtain the inequality
\be 
\label{eq:bineq}
\hspace{-0.1cm} B_{14} \gtrsim 0.15 \left( \frac{ M_{1.4}^{3/14} \dot{M}_{10}^{3/7}} {R_{6}^{15/14} \zeta_{-1}^{1/2} \xi^{3/2}} \right)^{14/5} \left( \frac{ \gamma_{\text{TEV}}} {2 \times 10^{6}} \right)^{14/5}.
\ee
{Using the propeller inversion \eqref{eq:mdotinv}, the above expression becomes
\begin{equation} \label{eq:bineq2}
B_{14} \gtrsim 1.4 L_{33}^{2} M_{1.4}^{-1} R_{6}^{-3} \zeta_{-1}^{-3} \xi^{-7}  \left( \frac{ \gamma_{\text{TEV}}} {2 \times 10^{6}}\right)^{6},
\end{equation}
which scales strongly with $\xi$.} The details of the requisite $B$ value depend on when the gamma-rays are observed relative to the orbital cycle.

\subsection{Orbital scenarios}

It has been argued by \cite{tor12} that the (non pulsational) radio emissions indicate that the highest energy emission occurs only at apastron, with 
lower, $\sim$GeV-band emissions at periastron.
If one assumes therefore that high-energy particles are only emitted at 
apastron where $\dot{M}$ is expected to achieve its local minimum, relatively weak fields could permit $\sim$TeV emission {for large enough $\xi$}. If we take $\xi \lesssim 0.3$ and fix the stellar parameters to their canonical values, requirement \eqref{eq:bineq2} becomes
\be \label{eq:bminz}
B_{p,\rm min} \sim 6 \times 10^{13} \zeta^{-3} (\xi/0.3)^{-7} L_{33}^{2} \text{ G}.
\ee
The above assumes an X-ray luminosity rate that is on the low end, as relevant for the source being at apastron, together with an extreme accelerator, $\zeta \sim 1$. Significantly larger values of 
$\dot{M} \sim 10^{-2} M_{\text{Edd}}$, as considered in Sec. \ref{sec:brake}, require polar field strengths that approach the virial limit unless 
$\xi \gg 0.3$. {Note that such values of $\xi$ are larger than those considered in Sec. \ref{sec:brake} by a factor $\gtrsim 2$, hinting again that the {maximum $\dot{P}$ is overinflated} unless the instantaneous $L_{\rm X}$ was much lower when the source pulsed in the radio.}

For context, we consider how this analysis matches up with other high-energy systems. Radio timing of the TeV binaries hosting the millisecond pulsars PSR J2032+4127 and PSR B1259--63, assuming standard dipole braking, gives $B_{p} \approx 3.4 \times 10^{12}$G and $B_{p} \approx 6.6 \times 10^{11}$G, respectively \citep{john94,cam09}. For the former source however, X-ray emissions are relatively weak, $L_{\rm X} \approx 1.1 \times 10^{31} \text{erg s}^{-1}$ \citep{cam09}, so that even the propeller-minimum \eqref{eq:bminz} imposes the modest constraint $B_{p,\rm min} \gtrsim 8 \times 10^{10} \zeta^{-7/5} (\xi/0.3)^{-7}$G to explain TeV outbursts. The X-ray luminosity of the latter is more like that of LSI, with $L_{\rm X} \sim 10^{33} \text{erg s}^{-1}$ at apastron \citep{kaw04}, though if $\xi \sim 1$ the required minimum is $B_{p, \rm min} \sim 1.4 \times 10^{11} \zeta^{-3}$G. The key difference for LSI is that we anticipate small values of $\xi$, if indeed $\dot{P} \gtrsim 10^{-11} \mbox{ ss}^{-1}$.

A key conclusion is that there is an inverse relationship between $\xi$, controlling the magnetospheric radius, and the polar field strength required to explain $\sim$~TeV emissions, at least within the context of the \cite{bed09b,bed09} model. This is the reverse of that imposed by the radio $\dot{P}$ constraints considered in Sec. \ref{sec:brake}; if $\xi$ is too small, synchrotron losses prevent the system from emitting high energy particles, while if it is too large the system cannot spindown fast enough. Overall, therefore, the analysis of this section suggests that polar field strengths of order $10^{13}$~G, at minimum, are required for LSI. If relativistic shocks dominate the acceleration process at the magnetospheric radius however \cite[i.e., if $\zeta \ll 1$;][]{khang07} then magnetar-level fields appear to be necessary.

%%%%%%%%%%%%%%%%%%%%%%%%%%%%%%%%%%%%%%%%%%

%%%%%%%%%%%%%%%%%%%%%%%%%%%%%%%%%%%%
\section{Radio-pulsar activation}
\label{sec:deathval}

It is generally put forth that pair production in charge-starved magnetospheric `gaps' is a necessary action in the powering of radio pulsations from neutron stars, magnetar or otherwise \citep{rs75} \cite[though cf.][]{melrose21}. Depending on the rotation rate, magnetic field, and the structure and location of these gaps, a variety of possible `death lines' arise, defining an overall `death valley' \citep{cr93,ha01,sz15}. This provides a basis for investigating the radio switch-on of LSI.

%%%%%%%%%%%%%%%%%%%%%%%
\subsection{Death valley}

Although many models exist, a promising location for these gaps is above the polar cap(s), as the open field line bundle ensures that material is continually expelled from there. The maximum potential drop, $\Delta V_{\text{max}}$, is
\be
 \label{eq:dvmax}
\Delta V_{\text{max}} \approx  2 \pi^2 \nu^2 B_{d} \Rs^3 / c^2
\ee
for \emph{dipole} component $B_{d}$. This must exceed the voltage drop $ \Delta V_{\rm pair} $ required for pair production; in a curvature-radiation scenario, we have \citep{rs75}
\be
\left( \frac {e \Delta V_{\text{pair}}} {m_{e} c^2} \right)^{3} \frac {\hbar} {2 m_{e} c R_{c}} \frac{H} {R_{c}} \frac{ B_{p}} {B_{\text{QED}}} \approx \frac {1}{15},
\ee
where $R_{c}$ is the curvature radius, $B_{\text{QED}} = m_{e}^2 c^3/ e \hbar \approx 4.4 \times 10^{13}$ G denotes the Schwinger field, $\hbar$ is the reduced Planck constant, and $H$ denotes a characteristic gap width. 

Following \cite{cr93}, we consider three models. (I) Pure dipole. Here one simply takes $R_{c} = (\Rs R_{\rm lc})^{1/2}$, $H = \Rs (\Rs / R_{\rm lc})^{1/2}$, and $B_{p} = B_{d}$. The death line, $\Delta V_{\text{max}} \geq \Delta V_{\text{pair}}$, reads $4 \log B_{d} - 7.5 \log P = 49.3$. (II) Twisted dipole. This case is similar to (I), though instead we assume that the magnetospheric field lines are curved such that $R_{c} \approx \Rs$. The line is given by $4 \log B_{d} - 6.5 \log P = 45.7$. (III) Twisted multipoles. In this case $B_{p} \gg B_{d}$ and $H \approx \Rs$ -- the field lines are so curved that radiated photons cross other parts of the cap's open field lines \cite[see Figure 3 in][]{cr93}. The line is given by $4 \log B_{d} - 6 \log P \approx 43.8$. It may be the case though that the cap is polluted by various materials, either from thermionic emissions from the crust \citep{sz15} or the accretion flow, in which case local electric fields may always be screened. So we also consider an outer-gap model, for which the death line (IV) reads $5 \log B_{d} - 12 \log P = 72$ [see Eq.~(24) of \cite{cr93}].

\begin{figure}
\includegraphics[width=\columnwidth]{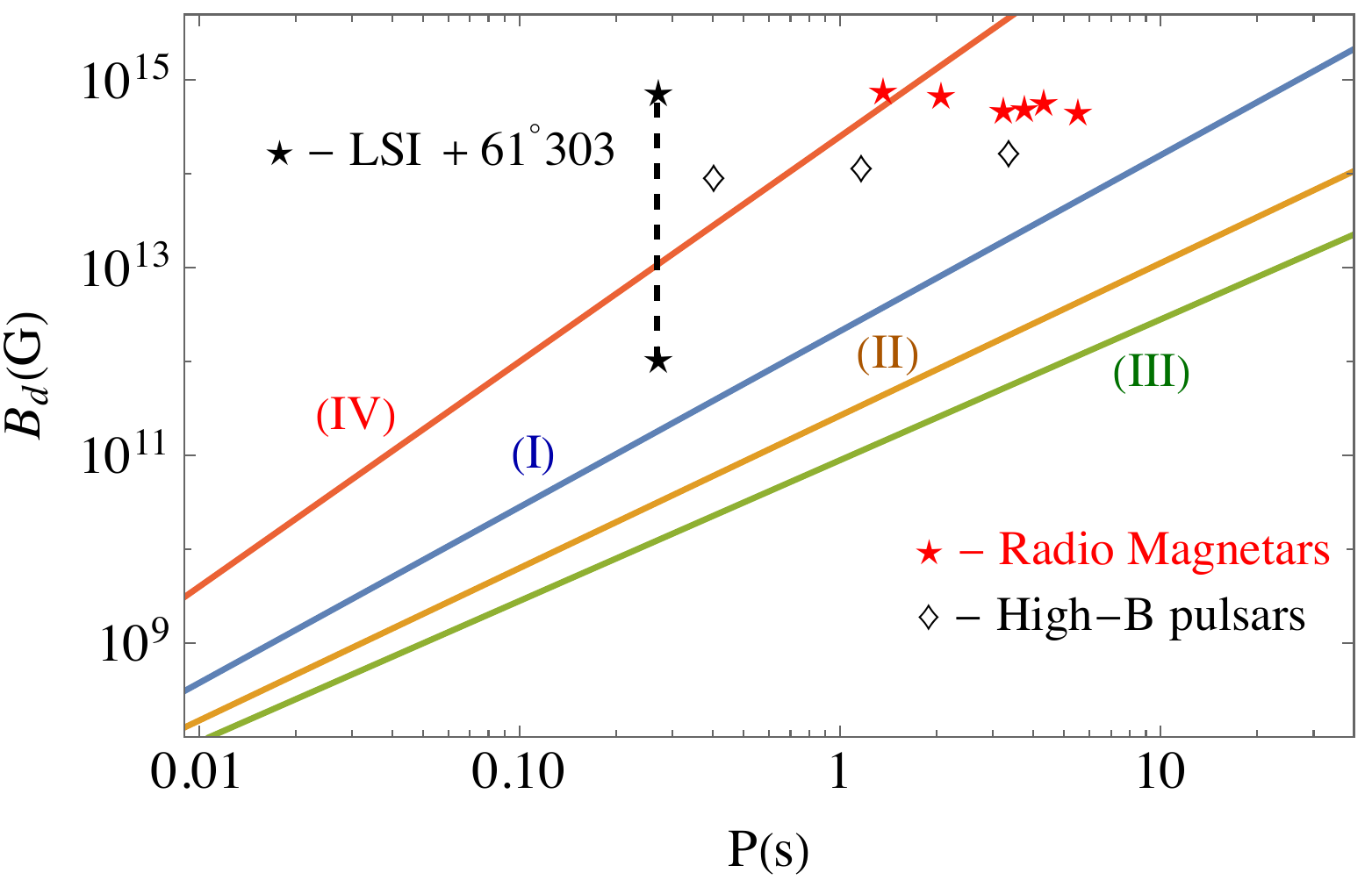}
\caption{Death lines (I--IV), detailed in the main text, in $B$--$P$ space. When a source lies vertically above a given line, the model says that the star can `switch on' as a radio pulsar. Known radio-loud magnetars (namely SGR J1745--2900, PSR J1622--4950, XTE J1810--197, 1E 1547.0--5408, {SGR 1935+2154, and Swift J1818.0--1607}) are shown by red stars, and a few high-B pulsars (PSR J1119--6127, PSR J1734--3333, and PSR J1718--3718) are shown with black diamonds. The primary in LSI is shown as a black star, with the dashed line delineating the possible $B$-field range. }
\label{fig:deathvalley}
\end{figure}

Figure \ref{fig:deathvalley} shows death lines I--IV on a $B$-$P$ diagram. The dashed line, headed by black stars, illustrates the possible range of field strengths for LSI. We also show for completeness the (characteristic) location of known radio-loud magnetars (red stars) and high-$B$ pulsars (black diamonds). It is clear that these latter sources can easily accommodate the death valley constraints, with the possible exception of the outer magnetosphere model (IV). Similarly, unless $B_{p} \lesssim 10^{13}$G for LSI it, too, easily survives the valley. 

{In a magnetar (i.e., strongly quantum electrodynamic) environment, photons may split with non-negligible probability, quenching pair production \citep{bar99}. \cite{med10} numerically traced the development of pair cascades in the magnetospheres of stars with $B \gtrsim 10^{14}$G, including effects from photon splitting with realistic selection rules for the polarisation modes, one-photon pair-production into low Landau levels, and resonant inverse Compton scattering from geometrically-diverse hotspots. Remarkably, they found that untwisted, dipole fields observe death lines similar to category (I) with $B_{d,\rm min} \approx 10^{12} \times (P/\text{s})^{2} \mbox{G}$ [see their Eq. (60)], which is $\lesssim 10^{11}$G for LSI.}

As such, conventional theory {and numerical simulations} suggest it is easy for LSI to switch on as a radio pulsar. So why are pulsations only now observed? {Aside from their $\sim \mu$Jy brightness --invisible prior to FAST -- dynamical phenomena related to the soft-gamma flares may be responsible. Patches of crust with intense magnetic spots can drift through a combination of plastic flow and the Hall effect, suggesting the emitting region may itself drift post-quake \citep{g22b}. Magnetospheric twist injections and diffusive settling could also reconfigure the polar cap geometry \citep{belo09}.}  Pollution of the gap region is also likely to play a role in the nulling fraction \citep{tor12}. Alternatively, the pulsar may be permanently `on' though have its radio emissions absorbed by the thick wind of the companion. {Future observations will help to settle the issue. If the pulsar is henceforth `off', at least prior to the release of another soft-gamma flare, this would point towards magnetospheric twist injections and diffusive reconfiguration. If pulses are consistently seen with orbitally-modulated brightnesses, a non-magnetar wind scenario, like that favoured for PSR B1259--63 \citep{john94}, may apply.}

%%%%%%%%%%%%%%%%%%%%%%%%%%
\section{Gravitational-wave observability}
\label{sec:GWs}

As explored in the previous sections, there are a number of inferences that can be made about the nature of LSI from the electromagnetic observations. 
Future pulse recordings from the source will be useful in narrowing down the {intrinsic} $\dot{P}$ values relative to $L_{\rm X}$, which seem to be the most 
important quantities as concerns the magnetar interpretation. If it turns out that LSI houses a sub-second magnetar,
the system could be visible in GWs. Having now a precise spin period for the object allows for targeted, narrow-band GW searches to be carried out, 
which can enhance the signal-to-noise by a factor $\sim \sqrt{T_{\text{obs}}}$ for observing time $T_{\text{obs}}$ \cite[e.g.,][]{watts08}. 

\subsection{Magnetic deformations}

Magnetic fields introduce
 mass-density asymmetries in a neutron star because the Lorentz force is not spherically symmetric \citep{cf53}. The inclusion of a toroidal field, 
which is necessary for the hydromagnetic stability of the system and {possibly} for powering the magnetar-like flares from the source, can also boost 
the effective deformation \citep{cut02,mast11,glamp18}. The resulting mass quadrupole moment varies in time as the star spins, resulting in the emission of 
GWs at a frequency of $f_{\text{GW}} = 2 \nu$. For LSI this implies that $f_{\text{GW}} = 7.43$Hz which, unfortunately, is difficult to observe with second-generation 
interferometers because the combination of seismic, gravitational-gradient (sometimes called `Newtonian'), and thermal noises severely limit detector sensitivity 
at $\lesssim 10$Hz \citep{ligo16}. This may not be a problem, however, for the future third-generation Einstein-Telescope (ET) which is designed to avoid these issues 
 by being underground and including cryogenic suspension technologies to reduce thermal vibrations \citep{ein10}.
 
Introducing the ellipticity $\epsilon = (I_{zz} - I_{xx})  / I_{\star}$, for moment of inertia tensor $I_{ij}$, the characteristic GW amplitude 
reads \cite[e.g.,][]{glamp18}
\be
\label{eq:charstr}
\begin{aligned}
h_{0} =& \frac {16 \pi^2 G \epsilon I_{\star} \nu^2} {c^4 d} \\
\approx& \, 2.2 \times 10^{-28} \left( \frac {\epsilon} {10^{-5}} \right) \left( \frac{\nu} {3.7 \text{ Hz}} \right)^{2} \left( \frac {2.65 \text{ kpc}} {d} \right).
\end{aligned}
\ee

The ellipticity depends critically on the relative contributions of the toroidal and poloidal field bulks.
To quantify its magnitude, we follow \cite{mast11} and introduce the quantity $0 < \Lambda \leq 1$, which denotes the ratio between the 
poloidal and the total energy; $\Lambda = 0.5$, for example, indicates equipartition between the internal toroidal and poloidal sectors. \cite{mast11} estimate the ellipticity as
\be
\label{eq:epsilon}
\epsilon \approx 6 \times 10^{-6} B_{p15}^2 R_{6}^{4} M_{1.4}^{-2} \left( 1 - { 0.389} / {\Lambda} \right).
\ee
Multipoles, proton superconductivity in the core, {or accreted mountains} can lead to increases in $\epsilon$.  Note the local minimum in $\epsilon$ at $\Lambda = 0.389$, which occurs because the prolate distortion induced by the toroidal field balances the poloidal, oblate distortion, leading to a force-free configuration.

\begin{figure}
\includegraphics[width=\columnwidth]{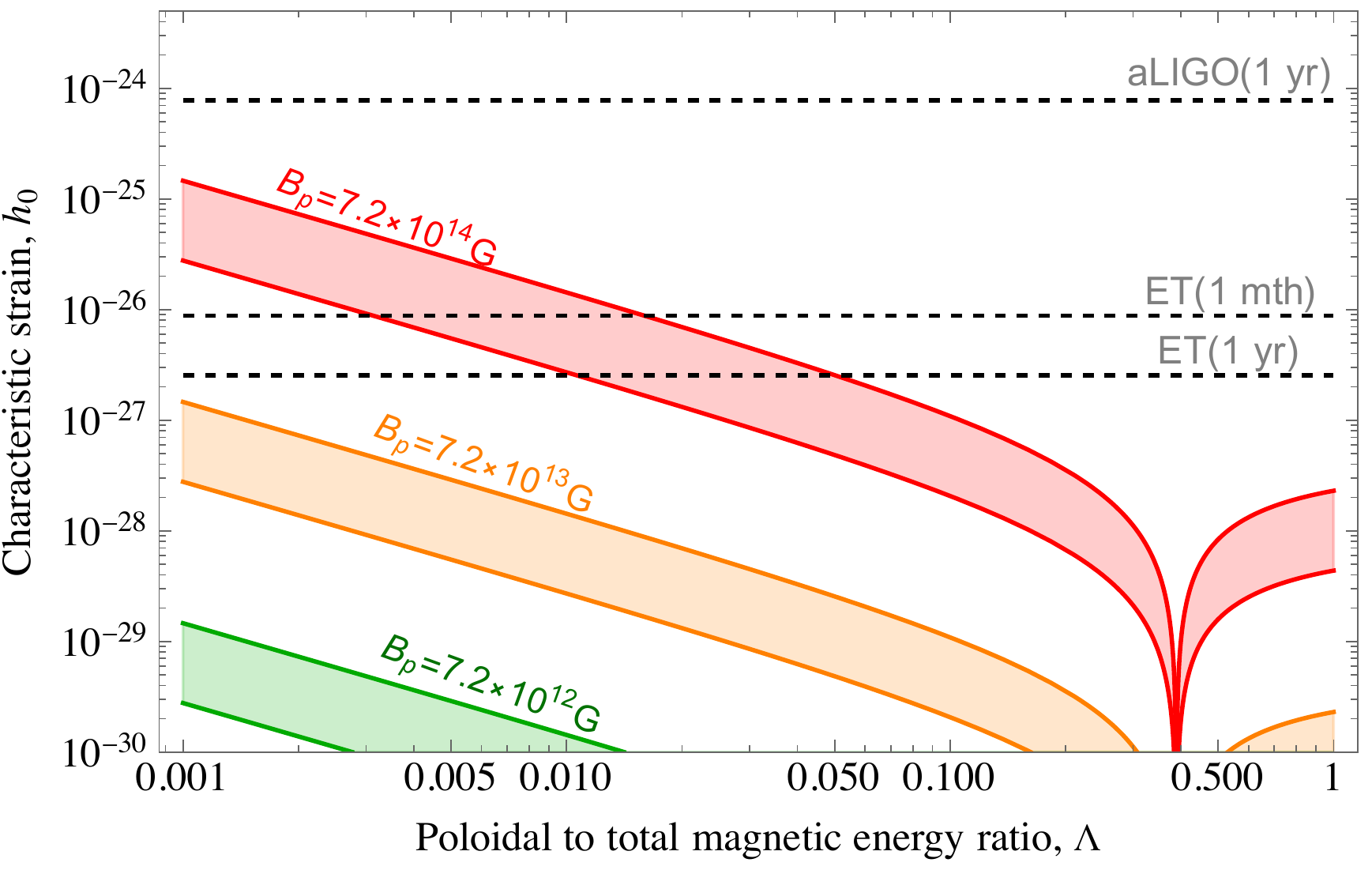}
\caption{Characteristic strains $h_{0}$, from Eq.~\eqref{eq:charstr}, as a function of the poloidal-to-total magnetic energy ratio, $\Lambda$, using the dipolar, 
Tolman-VII model \eqref{eq:epsilon} of \protect\cite{mast11} for LSI. The coloured bands represent the prediction for different $B_{p}$, with the lower curves having 
$M_{\star} = 1.4 M_{\odot}$ and $R_{\star} = 10$~km, and the upper curves having $M_{\star} = 2 M_{\odot}$ and $R_{\star} = 14$~km. Overlaid are the detection thresholds for 
aLIGO and ET for a variety of different observation times $T_{\text{obs}}$, as indicated by the labelled, dashed lines. %}
}
\label{fig:lsigws}
\end{figure}

Fig.~\ref{fig:lsigws} shows the characteristic strain \eqref{eq:charstr}, as a function of the poloidal-to-total energy ratio 
$\Lambda$ through \eqref{eq:epsilon}, for various field strengths $B_{p}$ (e.g., the orange bands correspond to $B_{p} = 7.2 \times 10^{13}$~G). 
In plotting expression \eqref{eq:charstr} for a given $B_{p}$, we consider both a canonical star with $M_{\star} = 1.4 M_{\odot}$ and $R_{\star} = 10$~km (lower curves) and an extreme one 
with $M_{\star} = 2 M_{\odot}$ and $R_{\star} = 14$~km (upper curves). Overlaid are relevant detection thresholds for 
aLIGO\footnote{Data from \url{https://dcc.ligo.org/public/0149/T1800044/005}}
and ET\footnote{Data from \url{http://www.et-gw.eu/index.php/etsensitivities}} 
for a few different observational times $T_{\text{obs}}$.
The figure demonstrates that the source is relatively dim in GWs, even for the extreme case $B_{p} = 7.2 \times 10^{14}$~G, as predicted if propeller torques are totally 
absent and the star spins down due to pure dipole braking (see Sec. \ref{sec:brake}). If, however, the star is relatively compact and contains a toroidal field that houses 
$\gtrsim 95\%$ of the total magnetic energy (i.e., $\Lambda \lesssim 0.05$), the system should be visible to ET given an observation time of $T_{\text{obs}} \gtrsim 6$ months. 
A superconducting core could allow for GWs to be detected by the star given a weaker toroidal field strength and fixed observation window \citep{cut02,land13}. 
These considerations strengthen the case for the construction of ET, as a detection of GWs from LSI would be able to significantly constrain its compactness and magnetic 
field strength. %which impact on each of the electromagnetic considerations detailed in previous sections.

%%%%%%%%%%%%%%%%%%%%%%
\section{Discussion}
\label{sec:conclusions}

We have provided a multi-faceted analysis of LSI, with the primary aim of deciphering the observational clues concerning its polar magnetic field strength. 
Although the $P, \dot{P}$ timing data (modulo the caveats noted in Sec. \ref{sec:sec2})
point to LSI being a  `typical' magnetar with $B_p \approx 7 \times 10^{14}\,\mbox{G}$,  other aspects of the system  suggest otherwise. The size of its magnetospheric radius and mass accretion rate (as estimated by the persistent
X-ray luminosity) imply that LSI probably experiences a strong propeller torque, leading to spindown. 
Attributing the bulk of a given $\dot{P}$ to a propeller torque implies that the system's true polar field instead lies in the ballpark 
of $\gtrsim 10^{13} \,\mbox{G}$ (see Fig. \ref{fig:propeller}), as is more typical for HMXBs. 

Other physical aspects of LSI lead to weaker constrains; a field of $B_p \gtrsim 10^{13} \,\mbox{G}$ places LSI well inside
the radio-loud portion of the magnetar `fundamental plane' of \cite{rea12}, and allows for the emission of $\sim$~TeV radiation from ultra-relativistic particles 
in relevant boundary layers \citep{bed09}. 
All together, our results suggest that LSI could be identified as a `low-$B$' magnetar or a `high-$B$' radio pulsar, 
the distinction between the two categories being, to some extent, a matter of semantics \citep{perna11}.

Future observations will shed more light on the true nature of LSI. On the electromagnetic side, one of the most critical parameters is $\dot{P}$, for which a long-term, timing-based solution is unavailable. Even if a lower $\dot{P}$ is recorded in future, which would imply a weaker polar field, this does not necessarily exclude the magnetar hypothesis however. The presence of multiple, short ($\sim 0.3$s) flares from LSI points towards the existence of an internal toroidal or {buried, multipolar poloidal} field. Indeed, simulations of magnetically-induced crust breaking in neutron stars, which are popular models for the progenitors of magnetar-like bursts \cite[e.g.,][]{perna11}, suggest that local fields of order $\gtrsim 10^{14}$~G are necessary to induce a fracture \citep{lan15}. Tighter constraints on a {possible} thermal component to the X-ray luminosity would be useful in this direction, as toroidal field decay is likely to provide a strong heat reservoir {(see Sec. \ref{sec:therm}).} A strong toroidal field also boosts the predicted ellipticity of the star, which provides a more optimistic outlook for GW detection (see Fig. \ref{fig:lsigws}). A confirmed detection
 of a continuous GW signal from LSI should be taken as strong evidence for the presence of 
an intense, $ B > 10^{14}\,\mbox{G}$, magnetic field, independently of the $\dot{P}$ value.

To date, there have been no confirmed detections of magnetars in binaries \cite[though see][for some candidates]{israel17,pop18}.
 It has been argued by \cite{kl19} that such systems must be rare. Notably, aside from the observations discussed in this article, there are formation considerations that we have not explored. For example, it may be the case that a large stockpile of angular momentum is required at birth to instigate substantial dynamo or magnetorotational activity, which suggests that (aside from merger possibilities) only rapidly-rotating stars undergoing core-collapse could produce magnetars. Whether any would-be companions can survive the supernova explosion in this case is unclear. The magnetic field may have decayed significantly therefore by the time the magnetar can find a companion, unless the decay process is stalled by plastic flows or strong crustal impurities, which extend the Ohmic timescale \citep{pop18}. Even so, further investigation into the theoretical and observational existence of magnetars in binaries is needed.

%%%%%%%%%%%%%%%%%%%%%%%%%%%%%%%%%%%%%%%%%%%
\section*{Acknowledgements}
AGS thanks Stefan Os{\l}owski for helpful discussions on pulse folding and general timing considerations.
KG acknowledges support from research grant PID2020-1149GB-I00 of the Spanish Ministerio de Ciencia e Innovaci{\'o}n. The research leading to these results has received funding from the
European Union's Horizon 2020 Programme under the AHEAD2020 project (grant n. 871158). {We are grateful to the anonymous referee for providing helpful and specific feedback that improved the quality of the manuscript.} 

%%%%%%%%%%%%%%%%%%%%%%%%%%%%%%%%%%%%%%%%%%%%%%%%%%%%%%%%%%%%%%%%
%%%%%%%%%%%%%%%%%%%%%%%%%%%%%%%%%%%%%%%%%%%%%%%%%%%%%%%%%%%%%%%%

\bibliographystyle{mn2e}

\begin{thebibliography}{94}
\expandafter\ifx\csname natexlab\endcsname\relax\def\natexlab#1{#1}\fi

\bibitem[{{Acciari} {et~al}\mbox{.}(2009){Acciari}, {Aliu}, {Arlen},
  {Bautista}, {Beilicke}, {Benbow}, {B{\"o}ttcher}, {Bradbury}, {Bugaev},
  {Butt}, {Butt}, {Byrum}, {Cannon}, {Cesarini}, {Chow}, {Ciupik}, {Cogan},
  {Colin}, {Cui}, {Daniel}, {Dickherber}, {Ergin}, {Falcone}, {Fegan},
  {Finley}, {Fortin}, {Fortson}, {Furniss}, {Gall}, {Gillanders}, {Grube},
  {Guenette}, {Gyuk}, {Hanna}, {Hays}, {Holder}, {Horan}, {Hui}, {Humensky},
  {Kaaret}, {Karlsson}, {Kieda}, {Kildea}, {Konopelko}, {Krawczynski},
  {Krennrich}, {Lang}, {LeBohec}, {Maier}, {McCann}, {McCutcheon}, {Millis},
  {Moriarty}, {Mukherjee}, {Nagai}, {Ong}, {Otte}, {Pandel}, {Perkins},
  {Perkins}, {Pohl}, {Quinn}, {Ragan}, {Reyes}, {Reynolds}, {Roache}, {Joachim
  Rose}, {Schroedter}, {Sembroski}, {Smith}, {Steele}, {Stroh}, {Swordy},
  {Theiling}, {Toner}, {Varlotta}, {Vassiliev}, {Wagner}, {Wakely}, {Ward},
  {Weekes}, {Weinstein}, {White}, {Williams}, {Wissel}, {Wood}, \&
  {Zitzer}}]{acc08}
{Acciari} V.~A. {et~al.}, 2009, \apj, 700, 1034

\bibitem[{{Aguilera}, {Pons} \& {Miralles}(2008){Aguilera}, {Pons}, \&
  {Miralles}}]{ag08}
{Aguilera} D.~N., {Pons} J.~A., {Miralles} J.~A., 2008, \apjl, 673, L167

\bibitem[{{Albert} {et~al}\mbox{.}(2006){Albert}, {Aliu}, {Anderhub},
  {Antoranz}, {Armada}, {Asensio}, {Baixeras}, {Barrio}, {Bartelt}, {Bartko},
  {Bastieri}, {Bavikadi}, {Bednarek}, {Berger}, {Bigongiari}, {Biland},
  {Bisesi}, {Bock}, {Bordas}, {Bosch-Ramon}, {Bretz}, {Britvitch}, {Camara},
  {Carmona}, {Chilingarian}, {Ciprini}, {Coarasa}, {Commichau}, {Contreras},
  {Cortina}, {Curtef}, {Danielyan}, {Dazzi}, {De Angelis}, {de los Reyes}, {De
  Lotto}, {Domingo-Santamar{\'\i}a}, {Dorner}, {Doro}, {Errando}, {Fagiolini},
  {Ferenc}, {Fern{\'a}ndez}, {Firpo}, {Flix}, {Fonseca}, {Font}, {Fuchs},
  {Galante}, {Garczarczyk}, {Gaug}, {Giller}, {Goebel}, {Hakobyan},
  {Hayashida}, {Hengstebeck}, {H{\"o}hne}, {Hose}, {Hsu}, {Isar}, {Jacon},
  {Kalekin}, {Kosyra}, {Kranich}, {Laatiaoui}, {Laille}, {Lenisa}, {Liebing},
  {Lindfors}, {Lombardi}, {Longo}, {L{\'o}pez}, {L{\'o}pez}, {Lorenz},
  {Lucarelli}, {Majumdar}, {Maneva}, {Mannheim}, {Mansutti}, {Mariotti},
  {Mart{\'\i}nez}, {Mase}, {Mazin}, {Merck}, {Meucci}, {Meyer}, {Miranda},
  {Mirzoyan}, {Mizobuchi}, {Moralejo}, {Nilsson}, {O{\~n}a-Wilhelmi},
  {Ordu{\~n}a}, {Otte}, {Oya}, {Paneque}, {Paoletti}, {Paredes}, {Pasanen},
  {Pascoli}, {Pauss}, {Pavel}, {Pegna}, {Persic}, {Peruzzo}, {Piccioli},
  {Poller}, {Pooley}, {Prandini}, {Raymers}, {Rhode}, {Rib{\'o}}, {Rico},
  {Riegel}, {Rissi}, {Robert}, {Romero}, {R{\"u}gamer}, {Saggion},
  {S{\'a}nchez}, {Sartori}, {Scalzotto}, {Scapin}, {Schmitt}, {Schweizer},
  {Shayduk}, {Shinozaki}, {Shore}, {Sidro}, {Sillanp{\"a}{\"a}}, {Sobczynska},
  {Stamerra}, {Stark}, {Takalo}, {Temnikov}, {Tescaro}, {Teshima}, {Tonello},
  {Torres}, {Torres}, {Turini}, {Vankov}, {Vitale}, {Wagner}, {Wibig},
  {Wittek}, {Zanin}, \& {Zapatero}}]{alb06}
{Albert} J. {et~al.}, 2006, Science, 312, 1771

\bibitem[{{Anzuini} {et~al}\mbox{.}(2022{\natexlab{a}}){Anzuini}, {Melatos},
  {Dehman}, {Vigan{\`o}}, \& {Pons}}]{anz22}
{Anzuini} F., {Melatos} A., {Dehman} C., {Vigan{\`o}} D., {Pons} J.~A.,
  2022{\natexlab{a}}, \mnras, 509, 2609

\bibitem[{{Anzuini} {et~al}\mbox{.}(2022{\natexlab{b}}){Anzuini}, {Melatos},
  {Dehman}, {Vigan{\`o}}, \& {Pons}}]{anz22b}
{Anzuini} F., {Melatos} A., {Dehman} C., {Vigan{\`o}} D., {Pons} J.~A.,
  2022{\natexlab{b}}, \mnras

\bibitem[{{Archibald} {et~al}\mbox{.}(2015){Archibald}, {Kaspi}, {Ng},
  {Scholz}, {Beardmore}, {Gehrels}, \& {Kennea}}]{arch15}
{Archibald} R.~F., {Kaspi} V.~M., {Ng} C.~Y., {Scholz} P., {Beardmore} A.~P.,
  {Gehrels} N., {Kennea} J.~A., 2015, \apj, 800, 33

\bibitem[{{Baring} \& {Harding}(1998)}]{bar99}
{Baring} M.~G., {Harding} A.~K., 1998, \apjl, 507, L55

\bibitem[{{Becker} {et~al}\mbox{.}(2012){Becker}, {Klochkov}, {Sch{\"o}nherr},
  {Nishimura}, {Ferrigno}, {Caballero}, {Kretschmar}, {Wolff}, {Wilms}, \&
  {Staubert}}]{beck12}
{Becker} P.~A. {et~al.}, 2012, \aap, 544, A123

\bibitem[{{Bednarek}(2009{\natexlab{a}})}]{bed09b}
{Bednarek} W., 2009{\natexlab{a}}, \aap, 495, 919

\bibitem[{{Bednarek}(2009{\natexlab{b}})}]{bed09}
{Bednarek} W., 2009{\natexlab{b}}, \mnras, 397, 1420

\bibitem[{{Beloborodov}(2009)}]{belo09}
{Beloborodov} A.~M., 2009, \apj, 703, 1044

\bibitem[{{Burrows} {et~al}\mbox{.}(2012){Burrows}, {Chester}, {D'Elia},
  {Palmer}, {Romano}, {Saxton}, {Sonbas}, {Stamatikos}, \& {Stratta}}]{bur12}
{Burrows} D.~N. {et~al.}, 2012, GRB Coordinates Network, 12914, 1

\bibitem[{{Camilo} {et~al}\mbox{.}(2009){Camilo}, {Ray}, {Ransom}, {Burgay},
  {Johnson}, {Kerr}, {Gotthelf}, {Halpern}, {Reynolds}, {Romani}, {Demorest},
  {Johnston}, {van Straten}, {Saz Parkinson}, {Ziegler}, {Dormody}, {Thompson},
  {Smith}, {Harding}, {Abdo}, {Crawford}, {Freire}, {Keith}, {Kramer},
  {Roberts}, {Weltevrede}, \& {Wood}}]{cam09}
{Camilo} F. {et~al.}, 2009, \apj, 705, 1

\bibitem[{{Casares} {et~al}\mbox{.}(2005){Casares}, {Ribas}, {Paredes},
  {Mart{\'\i}}, \& {Allende Prieto}}]{cas05}
{Casares} J., {Ribas} I., {Paredes} J.~M., {Mart{\'\i}} J., {Allende Prieto}
  C., 2005, \mnras, 360, 1105

\bibitem[{{Chandrasekhar} \& {Fermi}(1953)}]{cf53}
{Chandrasekhar} S., {Fermi} E., 1953, \apj, 118, 116

\bibitem[{{Chen} \& {Ruderman}(1993)}]{cr93}
{Chen} K., {Ruderman} M., 1993, \apj, 402, 264

\bibitem[{{Chernyakova} \& {Malyshev}(2020)}]{chern20}
{Chernyakova} M., {Malyshev} D., 2020, in Multifrequency Behaviour of High
  Energy Cosmic Sources - XIII. 3-8 June 2019. Palermo, p.~45

\bibitem[{{Cutler}(2002)}]{cut02}
{Cutler} C., 2002, \prd, 66, 084025

\bibitem[{{Das}, {Porth} \& {Watts}(2022){Das}, {Porth}, \& {Watts}}]{watts22}
{Das} P., {Porth} O., {Watts} A., 2022, arXiv e-prints, arXiv:2204.00249

\bibitem[{{Dubus}(2006)}]{dub06}
{Dubus} G., 2006, \aap, 456, 801

\bibitem[{{Dubus}(2013)}]{dub13}
{Dubus} G., 2013, \aapr, 21, 64

\bibitem[{{Dubus} \& {Giebels}(2008)}]{dub08}
{Dubus} G., {Giebels} B., 2008, The Astronomer's Telegram, 1715, 1

\bibitem[{{Esposito} {et~al}\mbox{.}(2007){Esposito}, {Caraveo}, {Pellizzoni},
  {de Luca}, {Gehrels}, \& {Marelli}}]{esp07}
{Esposito} P., {Caraveo} P.~A., {Pellizzoni} A., {de Luca} A., {Gehrels} N.,
  {Marelli} M.~A., 2007, \aap, 474, 575

\bibitem[{{Frail} \& {Hjellming}(1991)}]{frail91}
{Frail} D.~A., {Hjellming} R.~M., 1991, \aj, 101, 2126

\bibitem[{{Frail}, {Seaquist} \& {Taylor}(1987){Frail}, {Seaquist}, \&
  {Taylor}}]{frail87}
{Frail} D.~A., {Seaquist} E.~R., {Taylor} A.~R., 1987, \aj, 93, 1506

\bibitem[{{Frank}, {King} \& {Raine}(2002){Frank}, {King}, \&
  {Raine}}]{accbook}
{Frank} J., {King} A., {Raine} D., 2002, Accretion Power in Astrophysics, 3rd
  edn. Cambridge University Press

\bibitem[{{Fujisawa}, {Shota} \& {Kojima}(2022){Fujisawa}, {Shota}, \&
  {Kojima}}]{fuji22}
{Fujisawa} K., {Shota} K., {Kojima} Y., 2022, \mnras

\bibitem[{{Gavriil}, {Kaspi} \& {Woods}(2004){Gavriil}, {Kaspi}, \&
  {Woods}}]{gav04}
{Gavriil} F.~P., {Kaspi} V.~M., {Woods} P.~M., 2004, \apj, 607, 959

\bibitem[{{Ghosh} \& {Lamb}(1979)}]{ghosh79b}
{Ghosh} P., {Lamb} F.~K., 1979, ApJ, 234, 296

\bibitem[{{Glampedakis} \& {Gualtieri}(2018)}]{glamp18}
{Glampedakis} K., {Gualtieri} L., 2018, in Astrophysics and Space Science
  Library, Vol. 457, Astrophysics and Space Science Library, {Rezzolla} L.,
  {Pizzochero} P., {Jones} D.~I., {Rea} N., {Vida{\~n}a} I., eds., p. 673

\bibitem[{{Glampedakis} \& {Suvorov}(2021)}]{gs21}
{Glampedakis} K., {Suvorov} A.~G., 2021, \mnras, 508, 2399

\bibitem[{{Goldreich} \& {Julian}(1969)}]{gj69}
{Goldreich} P., {Julian} W.~H., 1969, \apj, 157, 869

\bibitem[{{Gourgouliatos}(2022)}]{g22b}
{Gourgouliatos} K.~N., 2022, arXiv e-prints, arXiv:2202.06662

\bibitem[{{Gourgouliatos} \& {Lander}(2021)}]{gl21}
{Gourgouliatos} K.~N., {Lander} S.~K., 2021, \mnras, 506, 3578

\bibitem[{{Granot} {et~al}\mbox{.}(2017){Granot}, {Gill}, {Younes}, {Gelfand},
  {Harding}, {Kouveliotou}, \& {Baring}}]{gran17}
{Granot} J., {Gill} R., {Younes} G., {Gelfand} J., {Harding} A., {Kouveliotou}
  C., {Baring} M.~G., 2017, \mnras, 464, 4895

\bibitem[{{Hadasch} {et~al}\mbox{.}(2012){Hadasch}, {Torres}, {Tanaka},
  {Corbet}, {Hill}, {Dubois}, {Dubus}, {Glanzman}, {Corbel}, {Li}, {Chen},
  {Zhang}, {Caliandro}, {Kerr}, {Richards}, {Max-Moerbeck}, {Readhead}, \&
  {Pooley}}]{had12}
{Hadasch} D. {et~al.}, 2012, \apj, 749, 54

\bibitem[{{Harding}, {Contopoulos} \& {Kazanas}(1999){Harding}, {Contopoulos},
  \& {Kazanas}}]{harding99}
{Harding} A.~K., {Contopoulos} I., {Kazanas} D., 1999, ApJL, 525, L125

\bibitem[{{Hardorp} {et~al}\mbox{.}(1959){Hardorp}, {Rohlfs}, {Slettebak}, \&
  {Stock}}]{hard59}
{Hardorp} J., {Rohlfs} K., {Slettebak} A., {Stock} J., 1959, Hamburger Sternw.
  Warner \& Swasey Obs., C01, 0

\bibitem[{{Hayasaki} \& {Okazaki}(2004)}]{hk04}
{Hayasaki} K., {Okazaki} A.~T., 2004, \mnras, 350, 971

\bibitem[{{Hibschman} \& {Arons}(2001)}]{ha01}
{Hibschman} J.~A., {Arons} J., 2001, \apj, 554, 624

\bibitem[{{Igoshev} \& {Popov}(2018)}]{pop18}
{Igoshev} A.~P., {Popov} S.~B., 2018, \mnras, 473, 3204

\bibitem[{{Israel} {et~al}\mbox{.}(2017){Israel}, {Belfiore}, {Stella},
  {Esposito}, {Casella}, {De Luca}, {Marelli}, {Papitto}, {Perri}, {Puccetti},
  {Castillo}, {Salvetti}, {Tiengo}, {Zampieri}, {D'Agostino}, {Greiner},
  {Haberl}, {Novara}, {Salvaterra}, {Turolla}, {Watson}, {Wilms}, \&
  {Wolter}}]{israel17}
{Israel} G.~L. {et~al.}, 2017, Science, 355, 817

\bibitem[{{Johnston} {et~al}\mbox{.}(1994){Johnston}, {Manchester}, {Lyne},
  {Nicastro}, \& {Spyromilio}}]{john94}
{Johnston} S., {Manchester} R.~N., {Lyne} A.~G., {Nicastro} L., {Spyromilio}
  J., 1994, \mnras, 268, 430

\bibitem[{{Kawachi} {et~al}\mbox{.}(2004){Kawachi}, {Naito}, {Patterson},
  {Edwards}, {Asahara}, {Bicknell}, {Clay}, {Enomoto}, {Gunji}, {Hara}, {Hara},
  {Hattori}, {Hayashi}, {Hayashi}, {Itoh}, {Kabuki}, {Kajino}, {Katagiri},
  {Kifune}, {Ksenofontov}, {Kubo}, {Kushida}, {Matsubara}, {Mizumoto}, {Mori},
  {Moro}, {Muraishi}, {Muraki}, {Nakase}, {Nishida}, {Nishijima}, {Ohishi},
  {Okumura}, {Protheroe}, {Sakurazawa}, {Swaby}, {Tanimori}, {Tokanai},
  {Tsuchiya}, {Tsunoo}, {Uchida}, {Watanabe}, {Watanabe}, {Yanagita},
  {Yoshida}, \& {Yoshikoshi}}]{kaw04}
{Kawachi} A. {et~al.}, 2004, \apj, 607, 949

\bibitem[{{Khangulyan} {et~al}\mbox{.}(2007){Khangulyan}, {Hnatic},
  {Aharonian}, \& {Bogovalov}}]{khang07}
{Khangulyan} D., {Hnatic} S., {Aharonian} F., {Bogovalov} S., 2007, \mnras,
  380, 320

\bibitem[{{King} \& {Lasota}(2019)}]{kl19}
{King} A., {Lasota} J.-P., 2019, \mnras, 485, 3588

\bibitem[{{Kravtsov} {et~al}\mbox{.}(2020){Kravtsov}, {Berdyugin}, {Piirola},
  {Kosenkov}, {Tsygankov}, {Chernyakova}, {Malyshev}, {Sakanoi}, {Kagitani},
  {Berdyugina}, \& {Poutanen}}]{krav20}
{Kravtsov} V. {et~al.}, 2020, \aap, 643, A170

\bibitem[{{Lander}(2013)}]{land13}
{Lander} S.~K., 2013, \prl, 110, 071101

\bibitem[{{Lander} {et~al}\mbox{.}(2015){Lander}, {Andersson}, {Antonopoulou},
  \& {Watts}}]{lan15}
{Lander} S.~K., {Andersson} N., {Antonopoulou} D., {Watts} A.~L., 2015, \mnras,
  449, 2047

\bibitem[{{Lattimer} {et~al}\mbox{.}(1991){Lattimer}, {Pethick}, {Prakash}, \&
  {Haensel}}]{page91}
{Lattimer} J.~M., {Pethick} C.~J., {Prakash} M., {Haensel} P., 1991, \prl, 66,
  2701

\bibitem[{{Lattimer} \& {Prakash}(2001)}]{lp01}
{Lattimer} J.~M., {Prakash} M., 2001, \apj, 550, 426

\bibitem[{{Li} {et~al}\mbox{.}(2011){Li}, {Torres}, {Zhang}, {Chen}, {Hadasch},
  {Ray}, {Kretschmar}, {Rea}, \& {Wang}}]{li11}
{Li} J. {et~al.}, 2011, \apj, 733, 89

\bibitem[{{Lindegren} {et~al}\mbox{.}(2021){Lindegren}, {Klioner},
  {Hern{\'a}ndez}, {Bombrun}, {Ramos-Lerate}, {Steidelm{\"u}ller}, {Bastian},
  {Biermann}, {de Torres}, {Gerlach}, {Geyer}, {Hilger}, {Hobbs}, {Lammers},
  {McMillan}, {Stephenson}, {Casta{\~n}eda}, {Davidson}, {Fabricius},
  {Gracia-Abril}, {Portell}, {Rowell}, {Teyssier}, {Torra}, {Bartolom{\'e}},
  {Clotet}, {Garralda}, {Gonz{\'a}lez-Vidal}, {Torra}, {Abbas}, {Altmann},
  {Anglada Varela}, {Balaguer-N{\'u}{\~n}ez}, {Balog}, {Barache}, {Becciani},
  {Bernet}, {Bertone}, {Bianchi}, {Bouquillon}, {Brown}, {Bucciarelli},
  {Busonero}, {Butkevich}, {Buzzi}, {Cancelliere}, {Carlucci}, {Charlot},
  {Cioni}, {Crosta}, {Crowley}, {del Peloso}, {del Pozo}, {Drimmel}, {Esquej},
  {Fienga}, {Fraile}, {Gai}, {Garcia-Reinaldos}, {Guerra}, {Hambly}, {Hauser},
  {Jan{\ss}en}, {Jordan}, {Kostrzewa-Rutkowska}, {Lattanzi}, {Liao}, {Licata},
  {Lister}, {L{\"o}ffler}, {Marchant}, {Masip}, {Mignard}, {Mints}, {Molina},
  {Mora}, {Morbidelli}, {Murphy}, {Pagani}, {Panuzzo}, {Pe{\~n}alosa Esteller},
  {Poggio}, {Re Fiorentin}, {Riva}, {Sagrist{\`a} Sell{\'e}s}, {Sanchez
  Gimenez}, {Sarasso}, {Sciacca}, {Siddiqui}, {Smart}, {Souami}, {Spagna},
  {Steele}, {Taris}, {Utrilla}, {van Reeven}, \& {Vecchiato}}]{gaia21}
{Lindegren} L. {et~al.}, 2021, \aap, 649, A2

\bibitem[{{Maraschi} \& {Treves}(1981)}]{mara81}
{Maraschi} L., {Treves} A., 1981, \mnras, 194, 1P

\bibitem[{{Martynov} {et~al}\mbox{.}(2016){Martynov}, {Hall}, {Abbott},
  {Abbott}, {Abbott}, {Adams}, {Adhikari}, {Anderson}, {Anderson}, {Arai},
  {Arain}, {Aston}, {Austin}, {Ballmer}, {Barbet}, {Barker}, {Barr},
  {Barsotti}, {Bartlett}, {Barton}, {Bartos}, {Batch}, {Bell}, {Belopolski},
  {Bergman}, {Betzwieser}, {Billingsley}, {Birch}, {Biscans}, {Biwer}, {Black},
  {Blair}, {Bogan}, {Bork}, {Bridges}, {Brooks}, {Celerier}, {Ciani}, {Clara},
  {Cook}, {Countryman}, {Cowart}, {Coyne}, {Cumming}, {Cunningham}, {Damjanic},
  {Dannenberg}, {Danzmann}, {Costa}, {Daw}, {DeBra}, {DeRosa}, {DeSalvo},
  {Dooley}, {Doravari}, {Driggers}, {Dwyer}, {Effler}, {Etzel}, {Evans},
  {Evans}, {Factourovich}, {Fair}, {Feldbaum}, {Fisher}, {Foley}, {Frede},
  {Fritschel}, {Frolov}, {Fulda}, {Fyffe}, {Galdi}, {Giaime}, {Giardina},
  {Gleason}, {Goetz}, {Gras}, {Gray}, {Greenhalgh}, {Grote}, {Guido}, {Gushwa},
  {Gustafson}, {Gustafson}, {Hammond}, {Hanks}, {Hanson}, {Hardwick}, {Harry},
  {Heefner}, {Heintze}, {Heptonstall}, {Hoak}, {Hough}, {Ivanov}, {Izumi},
  {Jacobson}, {James}, {Jones}, {Kandhasamy}, {Karki}, {Kasprzack}, {Kaufer},
  {Kawabe}, {Kells}, {Kijbunchoo}, {King}, {King}, {Kinzel}, {Kissel},
  {Kokeyama}, {Korth}, {Kuehn}, {Kwee}, {Landry}, {Lantz}, {Le Roux}, {Levine},
  {Lewis}, {Lhuillier}, {Lockerbie}, {Lormand}, {Lubinski}, {Lundgren},
  {MacDonald}, {MacInnis}, {Macleod}, {Mageswaran}, {Mailand}, {M{\'a}rka},
  {M{\'a}rka}, {Markosyan}, {Maros}, {Martin}, {Martin}, {Marx}, {Mason},
  {Massinger}, {Matichard}, {Mavalvala}, {McCarthy}, {McClelland}, {McCormick},
  {McIntyre}, {McIver}, {Merilh}, {Meyer}, {Meyers}, {Miller}, {Mittleman},
  {Moreno}, {Mueller}, {Mueller}, {Mullavey}, {Munch}, {Nuttall}, {Oberling},
  {O'Dell}, {Oppermann}, {Oram}, {O'Reilly}, {Osthelder}, {Ottaway},
  {Overmier}, {Palamos}, {Paris}, {Parker}, {Patrick}, {Pele}, {Penn},
  {Phelps}, {Pickenpack}, {Pierro}, {Pinto}, {Poeld}, {Principe}, {Prokhorov},
  {Puncken}, {Quetschke}, {Quintero}, {Raab}, {Radkins}, {Raffai}, {Ramet},
  {Reed}, {Reid}, {Reitze}, {Robertson}, {Rollins}, {Roma}, {Romie}, {Rowan},
  {Ryan}, {Sadecki}, {Sanchez}, {Sandberg}, {Sannibale}, {Savage}, {Schofield},
  {Schultz}, {Schwinberg}, {Sellers}, {Sevigny}, {Shaddock}, {Shao}, {Shapiro},
  {Shawhan}, {Shoemaker}, {Sigg}, {Slagmolen}, {Smith}, {Smith},
  {Smith-Lefebvre}, {Sorazu}, {Staley}, {Stein}, {Stochino}, {Strain},
  {Taylor}, {Thomas}, {Thomas}, {Thorne}, {Thrane}, {Torrie}, {Traylor},
  {Vajente}, {Valdes}, {van Veggel}, {Vargas}, {Vecchio}, {Veitch},
  {Venkateswara}, {Vo}, {Vorvick}, {Waldman}, {Walker}, {Ward}, {Warner},
  {Weaver}, {Weiss}, {Welborn}, {We{\ss}els}, {Wilkinson}, {Willems},
  {Williams}, {Willke}, {Winkelmann}, {Wipf}, {Worden}, {Wu}, {Yamamoto},
  {Yancey}, {Yu}, {Zhang}, {Zucker}, \& {Zweizig}}]{ligo16}
{Martynov} D.~V. {et~al.}, 2016, \prd, 93, 112004

\bibitem[{{Mastrano} {et~al}\mbox{.}(2011){Mastrano}, {Melatos}, {Reisenegger},
  \& {Akg{\"u}n}}]{mast11}
{Mastrano} A., {Melatos} A., {Reisenegger} A., {Akg{\"u}n} T., 2011, \mnras,
  417, 2288

\bibitem[{{Medin} \& {Lai}(2010)}]{med10}
{Medin} Z., {Lai} D., 2010, \mnras, 406, 1379

\bibitem[{{Melatos}(1997)}]{mel97}
{Melatos} A., 1997, \mnras, 288, 1049

\bibitem[{{Melrose}, {Rafat} \& {Mastrano}(2021){Melrose}, {Rafat}, \&
  {Mastrano}}]{melrose21}
{Melrose} D.~B., {Rafat} M.~Z., {Mastrano} A., 2021, \mnras, 500, 4530

\bibitem[{{Mirabel}, {Rodrigues} \& {Liu}(2004){Mirabel}, {Rodrigues}, \&
  {Liu}}]{mir04}
{Mirabel} I.~F., {Rodrigues} I., {Liu} Q.~Z., 2004, \aap, 422, L29

\bibitem[{{Mu{\~n}oz-Arjonilla} {et~al}\mbox{.}(2009){Mu{\~n}oz-Arjonilla},
  {Mart{\'\i}}, {Combi}, {Luque-Escamilla}, {S{\'a}nchez-Sutil}, {Zabalza}, \&
  {Paredes}}]{mun07}
{Mu{\~n}oz-Arjonilla} A.~J., {Mart{\'\i}} J., {Combi} J.~A., {Luque-Escamilla}
  P., {S{\'a}nchez-Sutil} J.~R., {Zabalza} V., {Paredes} J.~M., 2009, \aap,
  497, 457

\bibitem[{{Olausen} \& {Kaspi}(2014)}]{mcg14}
{Olausen} S.~A., {Kaspi} V.~M., 2014, \apjs, 212, 6

\bibitem[{{Papitto} {et~al}\mbox{.}(2013){Papitto}, {Ferrigno}, {Bozzo}, {Rea},
  {Pavan}, {Burderi}, {Burgay}, {Campana}, {di Salvo}, {Falanga},
  {Filipovi{\'c}}, {Freire}, {Hessels}, {Possenti}, {Ransom}, {Riggio},
  {Romano}, {Sarkissian}, {Stairs}, {Stella}, {Torres}, {Wieringa}, \&
  {Wong}}]{pap13}
{Papitto} A. {et~al.}, 2013, \nat, 501, 517

\bibitem[{{Papitto}, {Torres} \& {Li}(2014){Papitto}, {Torres}, \&
  {Li}}]{pap14}
{Papitto} A., {Torres} D.~F., {Li} J., 2014, \mnras, 438, 2105

\bibitem[{{Papitto}, {Torres} \& {Rea}(2012){Papitto}, {Torres}, \&
  {Rea}}]{pap12}
{Papitto} A., {Torres} D.~F., {Rea} N., 2012, \apj, 756, 188

\bibitem[{{Paredes} {et~al}\mbox{.}(2007){Paredes}, {Rib{\'o}}, {Bosch-Ramon},
  {West}, {Butt}, {Torres}, \& {Mart{\'\i}}}]{paer07}
{Paredes} J.~M., {Rib{\'o}} M., {Bosch-Ramon} V., {West} J.~R., {Butt} Y.~M.,
  {Torres} D.~F., {Mart{\'\i}} J., 2007, \apjl, 664, L39

\bibitem[{{Parfrey}, {Beloborodov} \& {Hui}(2013){Parfrey}, {Beloborodov}, \&
  {Hui}}]{parfey13}
{Parfrey} K., {Beloborodov} A.~M., {Hui} L., 2013, \apj, 774, 92

\bibitem[{{P{\'e}tri}(2019)}]{pet19}
{P{\'e}tri} J., 2019, \mnras, 485, 4573

\bibitem[{{Pons} \& {Perna}(2011)}]{perna11}
{Pons} J.~A., {Perna} R., 2011, \apj, 741, 123

\bibitem[{{Popov}(2016)}]{pop16}
{Popov} S.~B., 2016, Astronomical and Astrophysical Transactions, 29, 183

\bibitem[{{Punturo} {et~al}\mbox{.}(2010){Punturo}, {Abernathy}, {Acernese},
  {Allen}, {Andersson}, {Arun}, {Barone}, {Barr}, {Barsuglia}, {Beker},
  {Beveridge}, {Birindelli}, {Bose}, {Bosi}, {Braccini}, {Bradaschia}, {Bulik},
  {Calloni}, {Cella}, {Chassande Mottin}, {Chelkowski}, {Chincarini}, {Clark},
  {Coccia}, {Colacino}, {Colas}, {Cumming}, {Cunningham}, {Cuoco},
  {Danilishin}, {Danzmann}, {De Luca}, {De Salvo}, {Dent}, {De Rosa}, {Di
  Fiore}, {Di Virgilio}, {Doets}, {Fafone}, {Falferi}, {Flaminio}, {Franc},
  {Frasconi}, {Freise}, {Fulda}, {Gair}, {Gemme}, {Gennai}, {Giazotto},
  {Glampedakis}, {Granata}, {Grote}, {Guidi}, {Hammond}, {Hannam}, {Harms},
  {Heinert}, {Hendry}, {Heng}, {Hennes}, {Hild}, {Hough}, {Husa}, {Huttner},
  {Jones}, {Khalili}, {Kokeyama}, {Kokkotas}, {Krishnan}, {Lorenzini},
  {L{\"u}ck}, {Majorana}, {Mandel}, {Mandic}, {Martin}, {Michel}, {Minenkov},
  {Morgado}, {Mosca}, {Mours}, {M{\"u}ller{\textendash}Ebhardt}, {Murray},
  {Nawrodt}, {Nelson}, {Oshaughnessy}, {Ott}, {Palomba}, {Paoli}, {Parguez},
  {Pasqualetti}, {Passaquieti}, {Passuello}, {Pinard}, {Poggiani}, {Popolizio},
  {Prato}, {Puppo}, {Rabeling}, {Rapagnani}, {Read}, {Regimbau}, {Rehbein},
  {Reid}, {Rezzolla}, {Ricci}, {Richard}, {Rocchi}, {Rowan}, {R{\"u}diger},
  {Sassolas}, {Sathyaprakash}, {Schnabel}, {Schwarz}, {Seidel}, {Sintes},
  {Somiya}, {Speirits}, {Strain}, {Strigin}, {Sutton}, {Tarabrin},
  {Th{\"u}ring}, {van den Brand}, {van Leewen}, {van Veggel}, {van den Broeck},
  {Vecchio}, {Veitch}, {Vetrano}, {Vicere}, {Vyatchanin}, {Willke}, {Woan},
  {Wolfango}, \& {Yamamoto}}]{ein10}
{Punturo} M. {et~al.}, 2010, Classical and Quantum Gravity, 27, 194002

\bibitem[{{Ransom}, {Eikenberry} \& {Middleditch}(2002){Ransom}, {Eikenberry},
  \& {Middleditch}}]{ran02}
{Ransom} S.~M., {Eikenberry} S.~S., {Middleditch} J., 2002, \aj, 124, 1788

\bibitem[{{Rea} {et~al}\mbox{.}(2013){Rea}, {Israel}, {Pons}, {Turolla},
  {Vigan{\`o}}, {Zane}, {Esposito}, {Perna}, {Papitto}, {Terreran}, {Tiengo},
  {Salvetti}, {Girart}, {Palau}, {Possenti}, {Burgay},
  {G{\"o}{\u{g}}{\"u}{\c{s}}}, {Caliandro}, {Kouveliotou}, {G{\"o}tz},
  {Mignani}, {Ratti}, \& {Stella}}]{rip13}
{Rea} N. {et~al.}, 2013, \apj, 770, 65

\bibitem[{{Rea} {et~al}\mbox{.}(2012){Rea}, {Pons}, {Torres}, \&
  {Turolla}}]{rea12}
{Rea} N., {Pons} J.~A., {Torres} D.~F., {Turolla} R., 2012, \apjl, 748, L12

\bibitem[{{Romero} {et~al}\mbox{.}(2007){Romero}, {Okazaki}, {Orellana}, \&
  {Owocki}}]{rom07}
{Romero} G.~E., {Okazaki} A.~T., {Orellana} M., {Owocki} S.~P., 2007, \aap,
  474, 15

\bibitem[{{Ruderman} \& {Sutherland}(1975)}]{rs75}
{Ruderman} M.~A., {Sutherland} P.~G., 1975, \apj, 196, 51

\bibitem[{{Shaw} {et~al}\mbox{.}(2022){Shaw}, {Stappers}, {Weltevrede},
  {Brook}, {Karastergiou}, {Jordan}, {Keith}, {Kramer}, \& {Lyne}}]{shaw22}
{Shaw} B. {et~al.}, 2022, \mnras

\bibitem[{{Spitkovsky}(2006)}]{spit06}
{Spitkovsky} A., 2006, \apjl, 648, L51

\bibitem[{{Suvorov} \& {Melatos}(2020)}]{suvm20}
{Suvorov} A.~G., {Melatos} A., 2020, \mnras, 499, 3243

\bibitem[{{Szary}, {Melikidze} \& {Gil}(2015){Szary}, {Melikidze}, \&
  {Gil}}]{sz15}
{Szary} A., {Melikidze} G.~I., {Gil} J., 2015, \apj, 800, 76

\bibitem[{{Tam} {et~al}\mbox{.}(2010){Tam}, {Hui}, {Huang}, {Kong}, {Takata},
  {Lin}, {Yang}, {Cheng}, \& {Taam}}]{tam10}
{Tam} P.~H.~T. {et~al.}, 2010, \apjl, 724, L207

\bibitem[{{Thompson} \& {Duncan}(1993)}]{td93}
{Thompson} C., {Duncan} R.~C., 1993, \apj, 408, 194

\bibitem[{{Thompson} \& {Duncan}(1996)}]{td96}
{Thompson} C., {Duncan} R.~C., 1996, \apj, 473, 322

\bibitem[{{Thompson} {et~al}\mbox{.}(2000){Thompson}, {Duncan}, {Woods},
  {Kouveliotou}, {Finger}, \& {van Paradijs}}]{thomp00}
{Thompson} C., {Duncan} R.~C., {Woods} P.~M., {Kouveliotou} C., {Finger} M.~H.,
  {van Paradijs} J., 2000, \apj, 543, 340

\bibitem[{{Tian} {et~al}\mbox{.}(2007){Tian}, {Li}, {Leahy}, \&
  {Wang}}]{tian07}
{Tian} W.~W., {Li} Z., {Leahy} D.~A., {Wang} Q.~D., 2007, \apjl, 657, L25

\bibitem[{{Torres} {et~al}\mbox{.}(2012){Torres}, {Rea}, {Esposito}, {Li},
  {Chen}, \& {Zhang}}]{tor12}
{Torres} D.~F., {Rea} N., {Esposito} P., {Li} J., {Chen} Y., {Zhang} S., 2012,
  \apj, 744, 106

\bibitem[{{Torres} {et~al}\mbox{.}(2010){Torres}, {Zhang}, {Li}, {Rea},
  {Caliandro}, {Hadasch}, {Chen}, {Wang}, \& {Ray}}]{tor10}
{Torres} D.~F. {et~al.}, 2010, \apjl, 719, L104

\bibitem[{{Vigelius} \& {Melatos}(2009)}]{vig09}
{Vigelius} M., {Melatos} A., 2009, \mnras, 395, 1985

\bibitem[{{Wang}(1995)}]{wang95}
{Wang} Y.-M., 1995, ApJL, 449, L153

\bibitem[{{Watts} {et~al}\mbox{.}(2008){Watts}, {Krishnan}, {Bildsten}, \&
  {Schutz}}]{watts08}
{Watts} A.~L., {Krishnan} B., {Bildsten} L., {Schutz} B.~F., 2008, \mnras, 389,
  839

\bibitem[{{Weng} {et~al}\mbox{.}(2022){Weng}, {Qian}, {Wang}, {Torres},
  {Papitto}, {Jiang}, {Xu}, {Li}, {Yan}, {Liu}, {Ge}, \& {Yuan}}]{weng22}
{Weng} S.-S. {et~al.}, 2022, Nature Astronomy

\bibitem[{{White} {et~al}\mbox{.}(2022){White}, {Burrows}, {Coleman}, \&
  {Vartanyan}}]{white22}
{White} C.~J., {Burrows} A., {Coleman} M. S.~B., {Vartanyan} D., 2022, \apj,
  926, 111

\bibitem[{{Zdziarski}, {Neronov} \& {Chernyakova}(2010){Zdziarski}, {Neronov},
  \& {Chernyakova}}]{zdz10}
{Zdziarski} A.~A., {Neronov} A., {Chernyakova} M., 2010, \mnras, 403, 1873

\bibitem[{{Zhang}, {Gil} \& {Dyks}(2007){Zhang}, {Gil}, \& {Dyks}}]{zhang07}
{Zhang} B., {Gil} J., {Dyks} J., 2007, \mnras, 374, 1103

\end{thebibliography}

%%%%%%%%%%%%%%%%%% APPENDIX %%%%%%%%%%%%%%%%%%%%%%%%%%%%%%%%%%%%%%%

%%%%%%%%%%%%%%%%%%%%%%%%%%%%%%%%%%%%%%%%%%%%%

\label{lastpage}

\end{document}